\documentclass[aps,pra,10pt,twocolumn,floatfix,longbibliography]{revtex4-2}
\usepackage[T1]{fontenc}
\usepackage[utf8]{inputenc}
\usepackage{amsmath,amssymb,amsthm,mathtools,mathrsfs,graphicx,booktabs,array}
\usepackage[compat=0.3]{yquant}
\graphicspath{{figures/}}
\usepackage{xurl}
\usepackage[hidelinks]{hyperref}
\hypersetup{pdftitle={Device-independent witnesses of genuine multipartite entanglement and nonlocality for GHZ states and Bell-pair trees},
pdfauthor={Josep Batle; Tomasz Rybotycki; Tomasz Bialecki; Piotr Gawron; Adam Bednorz}}
\newtheorem{theorem}{Theorem}
\newtheorem{lemma}{Lemma}
\newtheorem{proposition}{Proposition}

\theoremstyle{definition}

\newcommand{\id}{1}
\newcommand{\ii}{\mathrm{i}}
\newcommand{\ket}[1]{\lvert #1\rangle}
\newcommand{\bra}[1]{\langle #1\rvert}
\newcommand{\avg}[1]{\langle #1\rangle}
\newcommand{\inner}[2]{\langle #1,#2\rangle}
\newcommand{\proj}[1]{\ket{#1}\!\bra{#1}}
\newcommand{\norm}[1]{\lVert #1\rVert}
\newcommand{\abs}[1]{\lvert #1\rvert}
\newcommand{\GHZ}{\mathrm{GHZ}}
\newcommand{\BS}{\mathrm{BS}}

\DeclareMathOperator{\Tr}{Tr}
\DeclareMathOperator{\Ree}{Re}

\DeclareMathOperator{\Var}{Var}
\DeclareMathOperator{\Cov}{Cov}
\newcommand{\doiurl}[1]{\href{https://doi.org/#1}{\nolinkurl{doi:#1}}}
\begin{document}
\title{Device-independent witnesses of genuine multipartite entanglement
and nonlocality for GHZ states and Bell-pair trees}
\author{Josep Batle$^{1}$}
\author{Tomasz Bia{\l}ecki$^{2,3}$}
\author{Adam Bednorz$^{2}$}
\email{abednorz@fuw.edu.pl}
\affiliation{$^1$CRISP -- Centre de Recerca Independent de sa Pobla, 07420 sa Pobla,
Balearic Islands, Spain}
\affiliation{$^2$Faculty of Physics, University of Warsaw,
ul. Pasteura 5, PL02-093 Warsaw, Poland}
\affiliation{$^3$Faculty of Physics and Applied Informatics, University of Lodz,
ul. Pomorska 149/153, PL90-236 Lodz, Poland}
\begin{abstract}
We derive and compare device-independent witnesses of genuine multipartite
entanglement (GME) and genuine multipartite nonlocality (GMNL) for GHZ states
and trees of Bell pairs. A binary-outcome family using four global measurement
contexts has the tight biseparable bound independent of the number of parties.
We distinguish these
GME bounds from Svetlichny-hybrid bounds, which permit arbitrary communication
within either side of a bipartition. For GHZ states, we also give a $2N$-context
GMNL witness and compare it with MABK and Svetlichny tests. For Bell-pair
trees, sums of CHSH and chained inequalities certify GME and GMNL,
respectively, with explicitly specified input protocols and multi-bit local
outputs. Separately, we prove the sharp four-context network bound
$I_\lambda\leq2$ for every $N\geq3$ and $\lambda\geq2$, and bound the MABK
scores of quantum networks assembled from bipartite sources. 
We successfully benchmark MABK, four-context, and Svetlichny witnesses on IBM Quantum
GHZ preparations up to $10$ parties, under the local-measurement and sampling assumptions. 
\end{abstract}
\maketitle

\section{Introduction}
A Bell-inequality violation rules out local hidden-variable descriptions and,
within quantum theory, certifies entanglement without requiring calibrated
measurement operators or known local dimensions \cite{bell,chsh,Brunner2014}.
For more than two laboratories, however, three questions must be separated:
whether the shared state is genuinely multipartite entangled, whether the
observed correlations are genuinely multipartite nonlocal, and whether the
resource can be assembled from bipartite quantum sources. These questions
have different null models and different thresholds.

The Mermin--Ardehali--Belinskii--Klyshko (MABK) family provides familiar
multipartite Bell tests \cite{mermin,ardehali,beli}. Its quantum-to-biseparable
maximal ratio is $\sqrt2$ \cite{tsirel} achieved by Greenberger-Horne-Zeilinger (GHZ) states \cite{ghz89,ghz90}, although the number of nonzero correlator contexts grows
exponentially. The GME threshold is higher, as it should exclude separable partitions 
\cite{seevinck01,nagata,uff,seevinck}
We derive a four-context alternative with a tunable penalty
for nonunanimous outcomes with the GME threshold is independent of $N$. 
For GMNL we use the stronger, explicitly stated
Svetlichny-hybrid null model \cite{svetlichny,seevinck,collinsm4,bancal11,Bancal2013}
and construct a separate $2N$-context GHZ witness.

On the other hand, a connected product of pure Bell pairs is GME with respect to its
\emph{laboratories}, even though every source has only two recipients 
\cite{coiteux,coiteux2,cao,mao,suprano,navasques,wolfe}.
Connected pure-state networks are known to permit GMNL under the definition
that requires no-signaling within each group \cite{Contreras2021}.
The Bell-tree construction below instead tests the larger Svetlichny-hybrid
class with unrestricted internal signaling. We give an explicit witness,
including its output alphabet and input schedule. We also
study the logically distinct problem of excluding bipartite-source network
preparations. Quantum Finner inequalities \cite{Finner1992,Renou2019} give
MABK and parameter-dependent four-context bounds. An operator-column
inequality proves the sharp triangle bound $I_2\leq2$; a monotone merging
argument then extends it to every $N\geq3$.

Experimental demonstration of GME of 6 qubits was acheived  device-independent \cite{barreiro}
while  24  qubits with  trusted  measurements  \cite{wang}.
On the other hand, 
IBM Quantum computers have been already used to test MABK and Svetlichny inequalities for up to 5 qubits
\cite{alsina,swain,gonzalez}.
 Here  we  present the test on IBM Quantum
which proved that GHZ states on up to 10 qubits  can be device-independent certified as  GME and GMNL 
up to trusted no-communication,  which cannot rely of the relatitistic speed constraints due to long time-scales  compared to  the qubit spacing.

After specifying the models, we present GHZ witnesses, network bounds,
Bell-tree witnesses, and the IBM Quantum data analysis. The analysis demonstrates up to 10-party
GME and GMNL without assumptions about the  Hilbert space operator representation, being a strong benchmark of the present capabilities of  quantum computers. Standard proofs and
longer optimizations are collected in the appendices.
The data and scripts are  publicly  available \cite{zen}.

\section{Models and notation}\label{sec:models}
\subsection{Measurements and device-independent assumptions}
There are $N$ spatially distinct or operationally isolated laboratories.
Except in the Bell-tree constructions, each has two binary measurements.
At laboratory $j$, the effects satisfy
\begin{equation}\label{eq:effects}
 A^{(j)}_{a|x}\succeq0,\qquad
 \sum_{a=\pm1}A^{(j)}_{a|x}=\id,\qquad
 A^{(j)}_x=A^{(j)}_{+|x}-A^{(j)}_{-|x}.
\end{equation}
Thus $A^{(j)}_x$ is a Hermitian contraction; it need not be a reflection.
Operators at different laboratories act on different tensor factors, and
$\avg{O}=\Tr\rho O$, for the state (Hermitian) $\rho\succeq 0$, $\Tr\rho=1$. 
In particular $P(a|x)=\langle A_{a|x}\rangle$ is a probability of the  output  $a$  for the  input $x$.
Products between laboratories denote tensor products, which result in the joint probability for effects.

Throughout, the plain symbol $1$ denotes the identity operator when used in
an operator expression, and an ordinary scalar otherwise; $c+O$ means
$c\,1+O$. The Hilbert-space support is indicated only when needed.
Pauli matrices, phase-dependent GHZ and Bell states, and the simultaneous
local reflection dilation are defined once in Appendix~\ref{app:notation}.
Pauli coordinates used inside a proof do not impose calibrated axes on the
devices.

A \emph{setting} is a local input, whereas a \emph{context} is a complete tuple
of local inputs. Input tuples may be correlated by the protocol, but their
realization must be independent of the preparation and pre-existing device
information. Each device receives only its prescribed local input. Outcomes
must be fixed without communication between laboratories. No outcome-based
filtering or postselection is allowed. These assumptions are not established
merely by observing a large score \cite{Brunner2014}.

\subsection{GME and GMNL}
A quantum state is biseparable if it admits a convex decomposition over
nontrivial bipartitions $L|\bar L$ of the form
\begin{equation}\label{eq:bisepstate}
 \rho=\sum_{L|\bar L,\eta}p_{L,\eta}\,
 \rho_L^{\eta}\otimes\rho_{\bar L}^{\eta},\qquad
 p_{L,\eta}\geq0,\quad\sum_{L,\eta}p_{L,\eta}=1.
\end{equation}
A state outside this convex set
is GME. The bipartition may change between components of a mixture.
For mixed states, entanglement across every fixed cut alone is not sufficient
to establish GME.

We use GMNL specifically for violation of the Svetlichny-hybrid model
of the input-dependent output probability,
\begin{equation}\label{eq:hybrid}
 P(\boldsymbol a|\boldsymbol x)=
 \sum_{L|\bar L,\eta}p_{L,\eta}
 P_L(\boldsymbol a_L|\boldsymbol x_L,\eta)
 P_{\bar L}(\boldsymbol a_{\bar L}|\boldsymbol x_{\bar L},\eta).
\end{equation}
Each group may use all its own inputs and arbitrary internal signaling, but
not the inputs of the other group. This is stronger than testing models
whose components are no-signaling within each group; the two definitions
must not be conflated \cite{Bancal2013}. Every biseparable quantum behavior
belongs to \eqref{eq:hybrid}, so a GMNL violation also certifies GME under the
quantum and isolation assumptions.

\subsection{Bipartite-source quantum networks}
For a loopless graph $G=(V,E)$, a fixed independent-source component has state
\begin{equation}\label{eq:networkcomponent}
 \Omega_\eta=\bigotimes_{e=\{u,v\}\in E}\rho_{uv}^{\eta},\qquad
 \mathcal H_u=\bigotimes_{v:\{u,v\}\in E}\mathcal H_{u|v}.
\end{equation}
Measurements at a vertex may act jointly on all incoming systems.
Local auxiliary states and trace-preserving local channels are allowed;
interlaboratory communication and postselection are not.
We also allow input-independent shared randomness $\eta$ in the source
states and local measurements. This gives the convex hull of the
independent-source behaviors. Missing edges can be represented by product
sources, so the complete graph covers all bipartite-source topologies.

Nonlinear inequalities apply to each independent-source component,
not in general to its shared-randomness average. A uniform \emph{linear}
score bound derived componentwise does extend to that average. We use this
distinction explicitly below. Excluding this network class is not synonymous
with certifying GME or GMNL: Bell-pair trees belong to the class but can
exhibit both properties.

\section{Witnesses for GHZ states}\label{sec:ghz}
\subsection{MABK normalization and biseparable bounds}\label{sec:mabk}
Introduce the non-Hermitian product operator
\begin{equation}\label{eq:Mcomplex}
 \mathcal M_N=\bigotimes_{j=1}^N(A^{(j)}_0+\ii A^{(j)}_1).
\end{equation}
Its real and imaginary parts collect, respectively, the even- and odd-weight
input strings, with coefficients $(-1)^{|\boldsymbol x|/2}$ and
$(-1)^{(|\boldsymbol x|-1)/2}$. Uffink's bound is
\begin{equation}\label{eq:uffink}
 \abs{\avg{\mathcal M_N}}\leq2^{N-1},\qquad N\geq2,
\end{equation}
whereas a single party has bound $\sqrt2$ \cite{uff}.
Appendix~\ref{app:standard} supplies a dimension-independent proof.
Define the normalized MABK \emph{score} \cite{mermin,ardehali,beli}, rather than an operator, by
\begin{equation}\label{eq:mabk}
 \mathcal B_N=2^{(1-N)/2}\Ree\!\left[
 e^{-\ii\vartheta_N}\avg{\mathcal M_N}\right],\qquad
 \vartheta_N=\frac{(N-1)\pi}{4}.
\end{equation}
For $N=2$ this is CHSH divided by $2$. Its sharp bounds are
\begin{equation}\label{eq:mabkbounds}
 \abs{\mathcal B_N}\leq
 \begin{cases}
 1,&\text{local},\\
 2^{(N-2)/2},&\text{biseparable},\quad N\geq3,\\
 2^{(N-1)/2},&\text{quantum}.
 \end{cases}
\end{equation}
The $N=2$ separable bound is $1$. The quantum maximum is attained with
$A^{(j)}_0=X$, $A^{(j)}_1=Y$, and
$\ket{\GHZ_N(\vartheta_N)}$ as defined in \eqref{eq:ghzdef}.
Equivalently, the phase can be absorbed into one party's measurement axes.

More generally, for a product partition containing $s$ singleton blocks and
$g\geq1$ blocks of size at least two, factorization and \eqref{eq:uffink} give \cite{nagata}
\begin{equation}\label{eq:partitionbound}
 \abs{\mathcal B_N}\leq2^{(N-s+1)/2-g}.
\end{equation}
Convexity extends this bound to mixtures of partitions of that type. The
all-singleton case must instead use the sharper local bound $1$.
There are $2^{N-1}$ nonzero MABK contexts when $N$ is odd and $2^N$ when $N$
is even.

\subsection{A four-context GME family}\label{sec:four}

The concept of the family below stems from  previous  inequalities
\cite{baccari,baccari2}
Distinguish one pivot laboratory $A$ and $m=N-1$ laboratories
$B^{(j)}$, $j\in\mathsf J:=\{2,\ldots,N\}$. Use only
\begin{equation}\label{eq:contexts}
 (x,y,\ldots,y),\qquad x,y\in\{0,1\}.
\end{equation}
The common input $y$ may be public as a synchronization rule, but its realized
value must not be sent to $A$, and $x$ must not be sent to any Bob before
outputs are fixed. Synchronization is not communication between outcomes.
Define
\begin{align}\label{eq:aggregate}
 R_0&=\prod_{j\in\mathsf J}B^{(j)}_0,&
 P_\pm&=\prod_{j\in\mathsf J}B^{(j)}_{\pm|1},\nonumber\\
 P&=P_++P_-,&\Gamma&=P_+-P_-,\qquad A^\pm=A_0\pm A_1.
\end{align}
The score is
\begin{equation}\label{eq:Ilambda}
 I_\lambda=\mathcal T-\lambda(1-p),\quad
 \mathcal T=\avg{A^+R_0}+\avg{A^-\Gamma},\quad p=\avg{P}.
\end{equation}
Here $p$ is the probability that all input-$1$ Bob outcomes agree.
The statistic $\Gamma$ assigns their common sign to unanimous outcomes and
zero otherwise. Every trial is retained; zero is a \emph{classically
processed score}, not a third local detector outcome.

\begin{theorem}[Four-context bounds]\label{thm:four}
For every $\lambda\geq0$, the local bound is $2$ and the quantum maximum is
$2\sqrt2$. For $N\geq3$ the tight biseparable bound is
\begin{equation}\label{eq:bsbound}
 b_\BS(\lambda)=
 \begin{cases}
 \dfrac{2\sqrt{\lambda^2+12}-\lambda}{3},&0\leq\lambda\leq2,\\[4pt]
 2,&\lambda\geq2.
 \end{cases}
\end{equation}
These bounds allow arbitrary local dimensions and binary POVMs.
\end{theorem}
The elementary local and quantum proofs are in
Appendix~\ref{app:fourbasic}; the complete tight biseparable optimization is in
Appendix~\ref{app:bisep}. In particular,
$I_0>4/\sqrt3$ and $I_2>2$ certify GME. For $N=2$, $p=1$ and every member is
CHSH, with separable bound $2$; the first branch of \eqref{eq:bsbound} does
not apply.

The mechanism at $\lambda=2$ is simple. For a product state across any cut,
one can relax the moments on the side not containing $A$ to obtain an
operator of the form
\begin{equation}\label{eq:shortK}
 K_2=A^+V+A^-Q-2(1-Q),
\end{equation}
where $V$ is a contraction and $Q$ a projection on the other part of Alice's
side. Since $2+A^-\succeq0$,
$K_2\preceq A^+V+A^-\preceq2$; the last inequality follows from
$|v+1|+|v-1|=2$ for $v\in[-1,1]$. The reduction, including cuts isolating
$A$, is detailed in Appendix~\ref{app:bisep}.

For $\ket{\GHZ_N(\phi)}$, choose
\begin{equation}\label{eq:fouraxes}
 B^{(j)}_0=X,\quad B^{(j)}_1=Z,\quad
 A_x=\tilde X_\alpha\cos\theta+(-1)^xZ\sin\theta.
\end{equation}
Using \eqref{eq:equatorial}, $p=1$ and
\begin{equation}\label{eq:fourghz}
 I_\lambda=2\cos\theta\cos(\phi-\alpha)+2\sin\theta.
\end{equation}
Thus $\alpha=\phi$ and $\theta=\pi/4$ attain $2\sqrt2$ for all $\lambda$.

This is a GME test, \emph{not} a Svetlichny-GMNL test. For $N\geq3$, a hybrid
block containing $A$ and one Bob learns both $x$ and $y$. It can choose
Alice's output to win all four CHSH-type terms while all Bobs output $+1$.
Hence its hybrid maximum is the algebraic value $4$ for every $\lambda\geq0$.
No quantum value of this score exceeds that hybrid bound.

\subsection{A signaling-sensitive CHSH illustration}\label{app:signaling}

It is instructive to see that  no-signaling is a  genuine restriction even for  bipartite Bell inequalities,
when enriched  with signaling-sensitive  terms \cite{almeida,kofler}.
Consider the signaling-sensitive CHSH functional
\begin{equation}\label{eq:signaling}
 \mathcal W=\sum_{x,y=0}^1\avg{(-1)^{xy}ab+(-1)^{x+y}(a+b)}_{xy}.
\end{equation}
Under no-signaling, the marginal terms cancel, so this is CHSH as in  Sec. \ref{sec:mabk}, with local,
quantum, and no-signaling bounds $2$, $2\sqrt2$, and $4$.
If only $A\to B$ signaling is allowed, deterministic responses are $a_x$
and $b_{xy}$. Alice's marginal terms cancel. Optimizing $b_{xy}$ gives
\begin{align}
 \max_b\mathcal W
 &=\sum_{x,y}|(-1)^{xy}a_x+(-1)^{x+y}|\nonumber\\
 &=2+2|a_1-1|\leq6.
\end{align}
The value $6$ requires $a_1=-1$. The opposite signaling direction has the
same bound by symmetry. With two-way signaling, the four independent
context maxima are $3,3,3,1$, giving $10$: at $x=y=1$, the expression
$-ab+a+b$ has maximum $1$.
This example shows why terms that vanish under no-signaling cannot simply
be removed when defining a signaling-hybrid null model.

\subsection{Svetlichny GMNL witness}\label{sec:svet}

For a bipartite separation, relaxing no-signaling within each separate  subset,
allows, like  above, a  larger families  of classical  strategies,  
limited  only by the lack of communication across the partition.
Use the same complex product \eqref{eq:Mcomplex} to define \cite{svetlichny}
\begin{equation}\label{eq:svet}
 \mathcal S_N=\Ree[(1+\ii)\avg{\mathcal M_N}].
\end{equation}
Its full correlator coefficients are $\pm1$, so all $2^N$ contexts are
required. The sharp bounds are
\begin{equation}\label{eq:svetbounds}
 \abs{\mathcal S_N}\leq
 \begin{cases}
 2^{\lfloor(N+1)/2\rfloor},&\text{local},\\
 2^{N-1},&\text{Svetlichny-hybrid},\\
 2^{N-1/2},&\text{quantum}.
 \end{cases}
\end{equation}
For $N\geq3$, $2^{N-1}$ is also the tight quantum-biseparable bound.
The quantum maximum uses $X,Y$ and $\ket{\GHZ_N(-\pi/4)}$.
These standard bounds \cite{svetlichny,seevinck,collinsm4} are proved in
Appendix~\ref{app:svet}. For even $N\geq4$, local relabelings make
$\mathcal B_N=2^{-N/2}\mathcal S_N$; their biseparable and
Svetlichny-hybrid thresholds then coincide. Thus an even-$N$ MABK score
above its biseparable threshold also certifies GMNL. For odd $N$, the
functionals differ: the MABK hybrid bound is $2^{(N-1)/2}$, equal to its
quantum maximum, so MABK alone cannot certify Svetlichny-GMNL.

\subsection{A \texorpdfstring{$2N$}{2N}-context GMNL witness}\label{sec:sparse}
Let $R_0$ be as in \eqref{eq:aggregate}, and let
\begin{equation}\label{eq:Rj}
 R_j=B^{(j)}_1\prod_{k\in\mathsf J\setminus\{j\}}B^{(k)}_0,
 \qquad j\in\mathsf J.
\end{equation}
Define
\begin{equation}\label{eq:KN}
 \mathcal K_N=\avg{A^+R_0}+\sum_{j\in\mathsf J}\avg{A^-R_j}.
\end{equation}
Only the all-zero Bob input and the $m$ single-changed-Bob inputs are used,
with either Alice input, for a total of $2(m+1)=2N$ contexts.

\begin{theorem}[Sparse GHZ GMNL witness]\label{thm:sparse}
The local and Svetlichny-hybrid bounds of \eqref{eq:KN} are both $2m$.
The quantum maximum is $2\sqrt{1+m^2}$, attained by GHZ states.
\end{theorem}
\begin{proof}
Fix a cut $\{A\}\cup U\mid V$ with $|U|=k$, $|V|=r=m-k\geq1$.
By linearity, group responses can be deterministic. Let $s_x=\pm1$ be the
parity of Alice's group at input $(x,0,\ldots,0)$, and let $t_0,t_j=\pm1$
be the other group's parities at its all-zero and single-changed inputs.
The baseline and the terms changed in $V$ contribute
\begin{equation}
 (s_0+s_1)t_0+(s_0-s_1)\sum_{j\in V}t_j\leq2r.
\end{equation}
Each of the $k$ remaining pairs of contexts contributes at most $2$.
Thus $\mathcal K_N\leq2r+2k=2m$ for every cut and mixture.
The local assignment $A_0=1$, $A_1=-1$, $B^{(j)}_y=1$ attains it.

For quantum strategies, put $Q_R=\sum_jR_j$.
The two-component Cauchy--Schwarz inequality, together with
$(A^+)^2+(A^-)^2\preceq4$, $\norm{R_0}\leq1$, and $\norm{Q_R}\leq m$, gives
\begin{equation}
 \abs{\mathcal K_N}\leq2\sqrt{1+m^2}.
\end{equation}
On $\ket{\GHZ_N(0)}$, choose $B^{(j)}_0=X$, $B^{(j)}_1=Y$ and
\begin{equation}\label{eq:sparseaxes}
 A_0=\frac{X-mY}{\sqrt{1+m^2}},\qquad
 A_1=\frac{X+mY}{\sqrt{1+m^2}}.
\end{equation}
The correlations $\avg{X^{\otimes N}}=1$ and
$\avg{Y_A Y_j\prod_{k\neq A,j}X_k}=-1$ give
$\mathcal K_N=2\sqrt{1+m^2}$.
\end{proof}
The quantum-to-hybrid ratio approaches one as $N$ increases. The context
saving therefore comes with a rapidly decreasing violation margin.

\section{Bounds for bipartite-source networks}\label{sec:networks}
\subsection{MABK and the unified Finner input}
\begin{theorem}[Network MABK bound]\label{thm:networkM}
Every bipartite-source quantum network, including its shared-randomness
convex hull, satisfies
\begin{equation}\label{eq:networkM}
 \abs{\avg{\mathcal M_N}}\leq2^{N/2},\qquad
 \abs{\mathcal B_N}\leq\sqrt2.
\end{equation}
\end{theorem}
Appendix~\ref{app:finner} proves this using a non-Hermitian tensor-contraction
bound; substituting $A_0+\ii A_1$ into an inequality stated only for positive
effects would not justify it. The bound is attainable with one Bell pair
and deterministic spectators. For $N\geq3$ it never exceeds the MABK
biseparable threshold, so MABK does not certify the GME of this network
class, although another witness can.

The same appendix collects the quantum Finner result needed here:
for independent bipartite sources and local effects $0\preceq E_j\preceq1$,
\begin{equation}\label{eq:finnermain}
 \avg{\bigotimes_jE_j}\leq\prod_j\avg{E_j}^{1/2}.
\end{equation}
Its positive-operator functional form and a covariance consequence are
also proved there. The exact triangle argument in
Appendix~\ref{app:triangle} instead uses an operator-column inequality.

\subsection{A parameter-dependent four-context envelope}\label{sec:netenvelope}
Fix one independent-source component and put
$p_+=\avg{P_+}$, $p_-=\avg{P_-}$, $p=p_++p_-$,
$u=p_+-p_-$, and $k=(N-1)/2$. For $N\geq3$, Finner with identity at $A$
and the arithmetic--geometric mean inequality yield
\begin{equation}\label{eq:finnerregion}
 p_+^{1/k}+p_-^{1/k}\leq1.
\end{equation}
Indeed, if $t_j=\avg{B^{(j)}_{+|1}}$, then
$p_+\leq\prod_j\sqrt{t_j}$ and
$p_-\leq\prod_j\sqrt{1-t_j}$.
The sum of their $1/k$ powers is at most
$\prod_jt_j^{1/(N-1)}+\prod_j(1-t_j)^{1/(N-1)}\leq1$.
This is a relaxation, not a characterization of network-realizable events.

Two independent quantum estimates are
\begin{equation}\label{eq:Tbounds}
 \mathcal T\leq2\sqrt{1+p},\qquad \mathcal T\leq2\sqrt{2-u^2}.
\end{equation}
The first follows after local dilation from $R_0^2=1$ and $\Gamma^2=P$;
the second is a CHSH marginal-bias bound \cite{pironio}, proved for the
present aggregate observables in Appendix~\ref{app:bias}.
Consequently a computable, not generally tight, network threshold is
\begin{equation}\label{eq:networkenvelope}
 \begin{split}
 I_\lambda&\leq b_{\mathrm{net}}^{\mathrm{env}}(N,\lambda)\\
 &:=\max_{\substack{p_+,p_-\geq0\\p_+^{1/k}+p_-^{1/k}\leq1}}
 \Bigl\{2\min\bigl[\sqrt{1+p},\sqrt{2-u^2}\bigr]\\
 &\hspace{43mm}-\lambda(1-p)\Bigr\}.
 \end{split}
\end{equation}
The maximum is taken \emph{before} averaging over shared randomness.
A uniform bound then extends by linearity of $I_\lambda$.

\subsection{The exact network bound for \texorpdfstring{$\lambda\geq2$}{lambda >= 2}}
We first give a short Finner proof for $N\geq5$, then use the exact
triangle result to cover every $N\geq3$.
\begin{theorem}[Finner bound for at least five parties]\label{thm:five}
For $N\geq5$, every bipartite-source network satisfies $I_2\leq2$.
The same holds for $\lambda\geq2$; the bound is attained by deterministic
outputs.
\end{theorem}
\begin{proof}
For $k\geq2$, \eqref{eq:finnerregion} implies
$\sqrt{p_+}+\sqrt{p_-}\leq1$. Hence
$4p_+p_-\leq(1-p)^2$ and $u^2\geq2p-1$.
By \eqref{eq:Tbounds},
\begin{equation}
 I_2\leq2\sqrt{3-2p}-2(1-p)\leq2,
\end{equation}
since $(2-p)^2-(3-2p)=(1-p)^2\geq0$.
The proof is componentwise and then linear under mixing.
Finally, $I_\lambda\leq I_2$ for $\lambda\geq2$.
\end{proof}

There is also a useful monotone reduction. For $N\geq4$, merge two
nonpivot parties $B,C$
into a laboratory $D$, with local observables
\begin{equation}\label{eq:merge}
 D_0=B_0C_0,\qquad D_1=(B_1+C_1)/2.
\end{equation}
The associated effects obey
\begin{equation}\label{eq:mergeeffect}
 D_{\pm|1}-B_{\pm|1}C_{\pm|1}
 =\frac{1-B_1C_1}{4}=:F\succeq0.
\end{equation}
This keeps agreeing signs and assigns a fair random sign on disagreement;
no trials are discarded. If $P_{\mathrm{rest},\pm}$ denote the other
unanimity effects, the score difference is
\begin{equation}\label{eq:mergepositive}
 \begin{split}
 I_2^{(N-1)}-I_2^{(N)}
 =\Bigl\langle F\bigl[&(2+A^-)P_{\mathrm{rest},+}\\
 &+(2-A^-)P_{\mathrm{rest},-}\bigr]\Bigr\rangle\geq0.
 \end{split}
\end{equation}
Indeed, all factors from different laboratories commute,
$F\succeq0$, $2\pm A^-\succeq0$, and
$P_{\mathrm{rest},\pm}\succeq0$.
An internal source becomes a local auxiliary state; parallel
external sources combine into one bipartite source. Thus merging preserves
the network class and reduces any $N\geq3$ instance to a triangle with
nondecreased score.

\begin{theorem}[Exact bipartite-source network bound]\label{thm:triangle}
For every $N\geq3$, every bipartite-source quantum network in
Sec.~\ref{sec:models}, including its shared-randomness convex hull, satisfies
\begin{equation}\label{eq:trianglebound}
 I_\lambda\leq2\qquad(\lambda\geq2).
\end{equation}
The bound is tight and allows arbitrary local binary POVMs and normal
trace-class source states on arbitrary Hilbert spaces.
\end{theorem}
\begin{proof}
Appendix~\ref{app:triangle} proves $I_2\leq2$ for an independent-source
triangle by an operator-column inequality and an exact nonnegative gap
identity. For each fixed independent-source component, repeated use of
\eqref{eq:mergepositive} reduces any $N\geq3$ network to a triangle without
decreasing its score. Local auxiliary states can be included in source
endpoints, and local trace-preserving channels can be absorbed into the
measurements through their unital positive adjoints. The resulting
measurements remain binary POVMs.

Since $I_2$ is linear in the observed probabilities, the same constant
bound holds after averaging over input-independent shared randomness.
Finally,
\begin{equation}
 I_\lambda=I_2-(\lambda-2)(1-p)\leq I_2\qquad(\lambda\geq2).
\end{equation}
Deterministic $+1$ outcomes at every laboratory give $p=1$ and
$I_\lambda=2$, establishing tightness.
\end{proof}
In particular, $I_2>2$ excludes the entire bipartite-source network class
for every $N\geq3$. The $N=2$ case is different: the score is CHSH, and a
single Bell source attains $I_2=2\sqrt2$.

\subsection{GME inside the network class: a two-Bell-pair example}\label{sec:star}
There are three laboratories and four qubits. Alice holds $a_B,a_C$, and
Bob and Charlie hold $b,c$, respectively. Use the Bell state of
\eqref{eq:belldef} to prepare
\begin{equation}\label{eq:starstate}
 \ket\Psi=\ket{\Phi^+}_{a_Bb}\otimes\ket{\Phi^+}_{a_Cc}
 =\frac12\sum_{r,s=0}^1\ket{rs}_A\ket r_B\ket s_C.
\end{equation}
Define Alice's even- and odd-parity flip operators
\begin{align}
 X_e&=\ket{00}\bra{11}+\ket{11}\bra{00},\nonumber\\
 Z_e&=\proj{00}-\proj{11},\qquad
 X_o=\ket{01}\bra{10}+\ket{10}\bra{01}.
\end{align}
With $B_0=C_0=X$, $B_1=C_1=Z$, and
\begin{equation}\label{eq:staraxes}
 A_0=(X_e+Z_e)/\sqrt2+X_o,\quad
 A_1=(X_e-Z_e)/\sqrt2+X_o,
\end{equation}
both $A_x^2=1$. Direct evaluation gives
\begin{equation}\label{eq:starvalue}
 p=\tfrac12,\qquad I_\lambda=1+\sqrt2-\tfrac\lambda2.
\end{equation}
In particular, $I_0=1+\sqrt2>4/\sqrt3$ certifies GME of this
bipartite-source state, without filtering, communication, or postselection.
It does not exclude the network model that generated the state.

\section{Witnesses for Bell-pair trees}\label{sec:trees}
Let $T=(V,E)$ be a tree with $|V|=N$ and $|E|=N-1$. The state
\begin{equation}\label{eq:treestate}
 \ket{\Psi_T}=\bigotimes_{e=\{u,v\}\in E}\ket{\Phi^+}_{u|v,\,v|u}
\end{equation}
uses $2(N-1)$ physical qubits. Laboratory $u$ holds $d_u$ qubits, one per
incident edge. Every nontrivial cut crosses an edge, so this pure state is
entangled across every cut and is GME with respect to the laboratories.
This statement does not identify laboratories with individual qubits.

For either witness below, a measurement at vertex $u$ returns a vector
$\boldsymbol a_u=(a_{u:e})_{e\ni u}\in\{\pm1\}^{d_u}$.
There are $2^{d_u}$ possible local outputs; thus this construction is
not a binary-outcome Bell test at vertices of degree greater than one. For a general local POVM
$\{E^{(u)}_{\boldsymbol a_u|x}\}$, the edge marginal observable is
\begin{equation}\label{eq:edgeobs}
 O^{(u)}_{e|x}=\sum_{\boldsymbol a_u}a_{u:e}
 E^{(u)}_{\boldsymbol a_u|x}.
\end{equation}
It is a contraction. The device-independent bounds do not assume that the
implemented POVM factorizes across incoming edge systems.

\subsection{Four-context GME from an edge-CHSH sum}\label{sec:treegme}
Two-color the tree, $V=U\sqcup W$. At each endpoint in $U$ use input $x=0,1$,
and at each endpoint in $W$ input $y=0,1$. The four synchronized contexts
measure every edge-CHSH score
\begin{align}\label{eq:treeCHSH}
 C_e={}&\avg{O^{(u)}_{e|0}O^{(v)}_{e|0}}
 +\avg{O^{(u)}_{e|1}O^{(v)}_{e|0}}\nonumber\\
 &+\avg{O^{(u)}_{e|0}O^{(v)}_{e|1}}
 -\avg{O^{(u)}_{e|1}O^{(v)}_{e|1}},\quad u\in U,\ v\in W,
\end{align}
and define $\mathcal C_T=\sum_{e\in E}C_e$.
\begin{theorem}[Tree GME witness]\label{thm:treegme}
The local, biseparable, and quantum bounds are, respectively,
\begin{equation}\label{eq:treegmebounds}
 2(N-1),\qquad 2\sqrt2(N-2)+2,\qquad2\sqrt2(N-1).
\end{equation}
All three are tight.
\end{theorem}
\begin{proof}
Across a fixed separable cut, every crossing edge has a separable endpoint
marginal and satisfies $C_e\leq2$. Every other edge satisfies
$C_e\leq2\sqrt2$. Since a tree cut crosses at least one edge, the middle
bound follows and is preserved under mixtures of cuts.
Removing one edge, putting a locally optimal product strategy on that edge,
and using Bell pairs on the others attains the biseparable bound.
The full Bell-tree state attains the quantum bound by measuring each
incident qubit in parallel, with $(X\pm Z)/\sqrt2$ at $U$ and $X,Z$ at $W$.
The all-local optimum is attained edge by edge with deterministic outputs.
\end{proof}
The biseparable loss from the quantum maximum is $2\sqrt2-2$.

\subsection{GMNL from chained edge tests: input protocol matters}\label{sec:treegmn}
For $L\geq2$ settings at each vertex, orient every edge from $U$ to $W$ and
use the chained score \cite{Braunstein1990,wehner}
\begin{align}\label{eq:chain}
 S_L^{(e)}={}&\sum_{i=1}^L\avg{O^{(u)}_{e|i}O^{(v)}_{e|i}}
 +\sum_{i=1}^{L-1}\avg{O^{(u)}_{e|i+1}O^{(v)}_{e|i}}\nonumber\\
 &-\avg{O^{(u)}_{e|1}O^{(v)}_{e|L}}.
\end{align}
For this GMNL test, \emph{each correlator is evaluated with all nonendpoint
laboratories held at input $1$}. Only the two endpoints of the tested edge
vary. The union of these edge-specific input tuples has at most
$2L(N-1)$ distinct global contexts; repeated tuples need only be measured
once. Every laboratory receives only its local input, not the edge label
or the full context. This is selection of measurement inputs, not
postselection of outcomes.

\begin{theorem}[Tree GMNL witness]\label{thm:treegmn}
With the preceding edge-specific protocol,
$\mathcal J_{T,L}=\sum_{e\in E}S_L^{(e)}$ satisfies
\begin{align}\label{eq:treegmnbounds}
 \mathcal J_{T,L}&\leq2L(N-1)-2&&\text{(Svetlichny-hybrid)},\nonumber\\
 \mathcal J_{T,L}&\leq2L(N-1)\cos\frac{\pi}{2L}&&\text{(quantum)}.
\end{align}
The quantum bound is attained by \eqref{eq:treestate}.
\end{theorem}
\begin{proof}
Fix a hybrid cut. For any crossing edge, all other inputs are fixed, so the
endpoint outputs factor across the cut into functions of the respective
endpoint input and shared randomness. That edge obeys the chained local
bound $2L-2$. Any internal edge has at most its algebraic value $2L$, even
with arbitrary within-group communication. At least one edge crosses,
which proves the hybrid bound and its extension to mixtures.
The dimension-independent quantum bound $2L\cos(\pi/2L)$ for each edge is
proved in Appendix~\ref{app:chain}. Bell pairs attain it simultaneously
using the product of the local edge measurements given there.
\end{proof}
The violation condition is
\begin{equation}\label{eq:treeviolation}
 4L(N-1)\sin^2\frac{\pi}{4L}<2.
\end{equation}
A simple sufficient choice is
$L>\pi^2(N-1)/8$, with integer $L\geq2$.
Thus every Bell-pair tree admits this GMNL certificate.

The four-context color synchronization used for GME cannot simply be
carried over to the signaling-hybrid proof. Already in a three-party star,
a hybrid group containing the center and one leaf learns both synchronized
inputs. It can then choose the center's two output bits to attain the
algebraic score on both edges, while the other leaf outputs a fixed sign.
The fixed-nonendpoint protocol above prevents that setting leakage.

\begin{table*}[t]
\caption{GHZ witness comparison for $N\geq3$, $m=N-1$. Bounds are in the
normalizations of the cited equations. The context counts refer to the
specified exact functionals, not to statistical shot complexity. The
four-context row uses $0\leq\lambda\leq2$.}
\label{tab:comparison}
\begin{ruledtabular}
\begin{tabular}{lllll}
Witness & Global contexts & Null model tested & Null threshold & Ideal GHZ value\\
\hline
$\mathcal B_N$, \eqref{eq:mabk} & $2^{N-1}$ odd $N$; $2^N$ even $N$ & Quantum biseparable & $2^{(N-2)/2}$ & $2^{(N-1)/2}$\\
$I_\lambda$, \eqref{eq:Ilambda} & $4$ & Quantum biseparable & $b_\BS(\lambda)$ & $2\sqrt2$\\
$\mathcal S_N$, \eqref{eq:svet} & $2^N$ & Svetlichny-hybrid & $2^{N-1}$ & $2^{N-1/2}$\\
$\mathcal K_N$, \eqref{eq:KN} & $2N$ & Svetlichny-hybrid & $2m$ & $2\sqrt{1+m^2}$
\end{tabular}
\end{ruledtabular}
\end{table*}

\section{GHZ witness benchmarks on IBM Quantum}\label{sec:experiment}
\subsection{Selected studies and physical implementation}

We have performed the above  tests to verify GME and GMNL of  GHZ  states up to 10 parties  on IBM Quantum  \cite{ibm},
using qubits  on \texttt{ibm\_kingston}: MABK and the four-context witness
 for $2\leq N\leq10$, and dedicated Svetlichny
runs at $N=3,5,7,9$. Each test corresponds to a single job. Even-$N$ Svetlichny results are obtained from the same MABK counts
by a fixed local relabeling, not from additional measurements. Each job
contains exactly $2^{21}=2\,097\,152$ shots, giving $46\,137\,344$ shots
in the selected data. The two-qubit runs are bipartite controls, not
multipartite certifications. No hardware data for the Bell-pair-tree
states of Sec.~\ref{sec:trees} are reported here.
The submitted options enable initialization, classified measurement, and
a $250\,\mu\mathrm{s}$ repetition delay; gate twirling, measurement
twirling, and dynamical decoupling are explicitly disabled.

\begin{figure*}[tbp]
\centering\includegraphics[width=.84\textwidth]{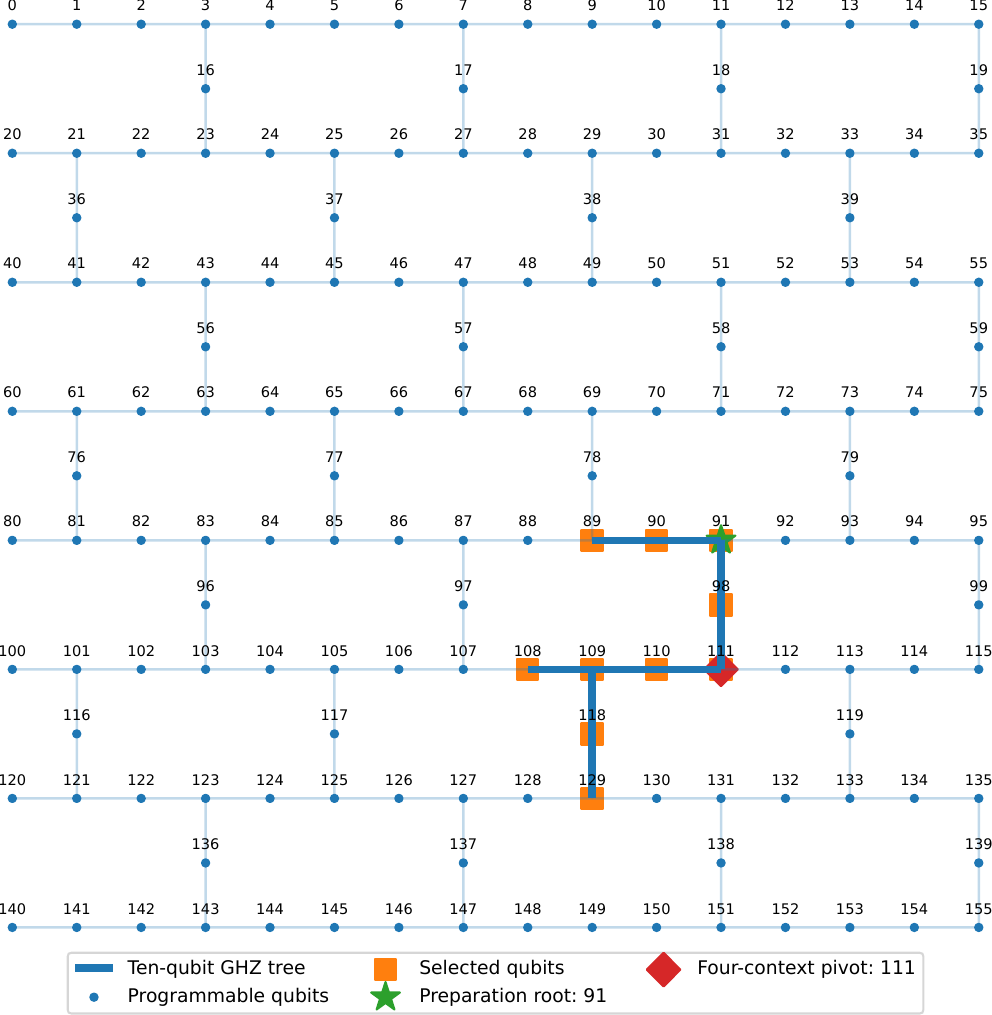}
\caption{Connectivity of \texttt{ibm\_kingston} and the frozen ten-qubit GHZ preparation
tree. Every label denotes a programmable qubit. Coordinates are a schematic embedding, not
measured chip coordinates. Thick edges mark the nine selected couplings,
with preparation root $91$ and four-context pivot $111$. }
\label{fig:kingston}
\end{figure*}

Kingston is a Heron r2 device with $156$ programmable qubits. 
Figure~\ref{fig:kingston} shows the $176$ undirected CZ
couplings in the saved experiment-time calibration snapshot. The highlighted
GHZ preparation uses physical qubits
\begin{equation}\label{eq:physical-tree}
 (89,90,91,98,111,110,109,118,108,129).
\end{equation}
The root is qubit $91$, and the distinguished four-context party is qubit
$111$. The recorded directed CNOT layers, in physical labels, are
\begin{align}\label{eq:physical-layers}
 &(91\!\to\!90),\quad(90\!\to\!89,\,91\!\to\!98),\notag\\
 &(98\!\to\!111),\quad(111\!\to\!110),\quad(110\!\to\!109),\notag\\
 &(109\!\to\!118),\quad(109\!\to\!108,\,118\!\to\!129).
\end{align}
Each CNOT is emitted as target Hadamard--CZ--target Hadamard. Thus the
$N$-qubit circuit uses $N-1$ CZ gates. Smaller frozen trees are obtained
by successively removing physical leaves $129,108,118,109,110,89,90$;
the $N=2$ endpoint is the pair $(98,111)$, with root $98$. The frozen
plans, not the geometric appearance of the figure, define the preparation
order. Previous tree-selection runs are not pooled into these measurements.
A GHZ state prepared along a tree is not a tensor product of Bell pairs
on that tree: the latter assigns multiple local qubits to internal vertices.

\subsection{From observables to native gates}
The experiments submit already physically mapped circuits using only
$Z_\vartheta$, $X_+$, $CZ$, a preparation-boundary barrier, and measurement.
Here $Z_\vartheta=\exp(-\ii\vartheta Z/2)$ and
$X_+=\exp(-\ii\pi X/4)$, as in Appendix~\ref{app:ibm-gates}; global
phases are irrelevant. There is no
Qiskit transpilation pass; fixed single-qubit Clifford blocks are folded
only within the preparation, never across the setting boundary. In time
order, the terminal block is
\begin{equation}\label{eq:native-measure}
 Z_{\pi/2-\alpha}\to X_+\to Z\text{-measurement}.
\end{equation}
Writing $V_\alpha=X_+ Z_{\pi/2-\alpha}$ gives
$V_\alpha^\dagger ZV_\alpha=X\cos\alpha+Y\sin\alpha$, with the Pauli
conventions of Appendix~\ref{app:notation}. Outcome bit $b$ represents
the sign $(-1)^b$; logical bit zero is the rightmost bit of each archived
count string. The fixed logical-to-physical mapping is saved in every plan.

For the four-context study, the preparation is
$F^{\otimes N}\ket{\GHZ_N(0)}$, where $F=X_-$. Since
$F^\dagger XF=X$ and $F^\dagger YF=Z$, choosing
$\alpha_A=(-1)^x\pi/4$ and $\alpha_{B_j}=y\pi/2$ realizes, in the GHZ
frame, precisely $A_x=(X+(-1)^xZ)/\sqrt2$, $B_{j0}=X$, and $B_{j1}=Z$.
The two settings therefore vary only local readout rotations. In MABK and
Svetlichny runs the $F$ rotations are omitted, a prescribed root phase is
inserted, and every party measures $X$ or $Y$. Appendix~\ref{app:ibm-gates}
gives the exact phase and outcome relabelings needed to match the paper's
polynomial conventions, together with representative
circuits. No phase fitting, sign maximization, or absolute-value selection
is performed on the measured data.

\subsection{Raw-count results and normalization}
We put the two full-correlation families on the common scale
\begin{equation}\label{eq:ibm-scale}
 W_N^{\rm M}=2^{(4-N)/2}\mathcal B_N,\qquad
 W_N^{\rm S}=\frac{2\mathcal S_N}{2^{N-1}}.
\end{equation}
Their relevant certification thresholds are $2$ and their quantum maxima
are $2\sqrt2$. For MABK the threshold means quantum biseparability
($N\geq3$); for Svetlichny it means the signaling-within-group hybrid
model. The fully local bounds are
$2^{2-N/2}$ and $2^{2-N+\lfloor(N+1)/2\rfloor}$, respectively. For the
four-context family, $2$ is the local bound for all reported penalties,
whereas its GME bound is $b_\BS(\lambda)$.

\begin{table}[t]
\caption{MABK raw-count results on the scale $W_N^{\rm M}=2^{(4-N)/2}\mathcal B_N$. The biseparable threshold is $2$ ($N\geq3$), the quantum maximum is $2\sqrt2$, and the fully local bound $b_{\rm L}$ depends on $N$. $K$ counts measured contexts. All uncertainties are one-standard-error shot-noise estimates. $N=2$ is a bipartite control.}
\label{tab:ibm-mabk}\centering\begin{ruledtabular}\begin{tabular}{rrrrr}
$N$ & $K$ & $W_N^{\rm M}$ & SE & $b_{\rm L}$\\
2 & 4 & 2.64306 & 0.00207 & 2.0000 \\
3 & 4 & 2.52144 & 0.00088 & 1.4142 \\
4 & 16 & 2.47525 & 0.00216 & 1.0000 \\
5 & 16 & 2.43235 & 0.00100 & 0.7071 \\
6 & 64 & 2.39911 & 0.00221 & 0.5000 \\
7 & 64 & 2.29230 & 0.00114 & 0.3536 \\
8 & 256 & 2.26537 & 0.00226 & 0.2500 \\
9 & 256 & 2.18574 & 0.00124 & 0.1768 \\
10 & 1024 & 2.09865 & 0.00231 & 0.1250 \\
\end{tabular}\end{ruledtabular}\end{table}

\begin{table*}[t]
\caption{Four-context scores recomputed at the three requested penalties from the same raw data. Errors include correlator--unanimity covariance. Each run has four contexts, $64$ complete setting sweeps, and $8192$ shots per circuit. The three columns are correlated reanalyses, not independent experiments. The recorded original primary penalty was $1/\sqrt2$, not $\sqrt2$; its results remain in the data release. $p$ is the mean of the two measured Bob-unanimity frequencies.}
\label{tab:ibm-four}\centering\begin{ruledtabular}\begin{tabular}{rcccc}
$N$ & $I_0\pm\mathrm{SE}$ & $I_{\sqrt2}\pm\mathrm{SE}$ & $\widehat I_2\pm\mathrm{SE}$ & $p$\\
2 & $2.64830\pm0.00207$ & $2.64830\pm0.00207$ & $2.64830\pm0.00207$ & 1.000000 \\
3 & $2.62319\pm0.00207$ & $2.59360\pm0.00209$ & $2.58134\pm0.00211$ & 0.979074 \\
4 & $2.56014\pm0.00207$ & $2.48532\pm0.00214$ & $2.45433\pm0.00218$ & 0.947098 \\
5 & $2.54274\pm0.00208$ & $2.46770\pm0.00215$ & $2.43662\pm0.00219$ & 0.946938 \\
6 & $2.46971\pm0.00209$ & $2.34713\pm0.00220$ & $2.29635\pm0.00226$ & 0.913321 \\
7 & $2.39925\pm0.00212$ & $2.24966\pm0.00224$ & $2.18770\pm0.00231$ & 0.894223 \\
8 & $2.36430\pm0.00212$ & $2.19861\pm0.00226$ & $2.12998\pm0.00234$ & 0.882843 \\
9 & $2.35120\pm0.00212$ & $2.17447\pm0.00226$ & $2.10126\pm0.00235$ & 0.875031 \\
10 & $2.30456\pm0.00214$ & $2.11237\pm0.00229$ & $2.03276\pm0.00238$ & 0.864100 \\
\end{tabular}\end{ruledtabular}\end{table*}

\begin{table}[t]
\caption{Svetlichny scores on the scale $W_N^{\rm S}=2\mathcal S_N/2^{N-1}$: hybrid threshold $2$, quantum maximum $2\sqrt2$. Even-$N$ entries are the very same counts and errors as Table~\ref{tab:ibm-mabk}, after a fixed local relabeling; they are not extra jobs. Odd-$N$ entries use the dedicated October study.}
\label{tab:ibm-svet}\centering\begin{ruledtabular}\begin{tabular}{rrrr}
$N$ & $W_N^{\rm S}$ & SE & Origin\\
3 & 2.50167 & 0.00216 & Dedicated \\
4 & 2.47525 & 0.00216 & MABK reuse \\
5 & 2.43186 & 0.00219 & Dedicated \\
6 & 2.39911 & 0.00221 & MABK reuse \\
7 & 2.28200 & 0.00227 & Dedicated \\
8 & 2.26537 & 0.00226 & MABK reuse \\
9 & 2.24017 & 0.00228 & Dedicated \\
10 & 2.09865 & 0.00231 & MABK reuse \\
\end{tabular}\end{ruledtabular}\end{table}

Tables~\ref{tab:ibm-mabk}--\ref{tab:ibm-svet} report estimates and one-standard-error
shot-noise uncertainties. The MABK support has $2^N$ contexts for even $N$
and $2^{N-1}$ for odd $N$; every supported context is measured. In MABK,
$N=2,3$ use $64$ complete setting sweeps, $N=4,5$ use $16$, $N=6,7$ use
$4$, and $N=8,9,10$ use one. The shots per circuit are $8192$, except
$N=10$, where they are $2048$. The dedicated Svetlichny runs use all
$2^N$ contexts, with $32,8,2,1$ sweeps at $N=3,5,7,9$ respectively;
the first three have $8192$ shots per circuit and $N=9$ has $4096$.
All four-context runs use $64$ sweeps of the four contexts, with $8192$
shots per circuit. A sweep here is a saved submission grouping, not
an independently verified chronological block or a per-shot random input.

\begin{figure}[tbp]
\includegraphics[width=\columnwidth]{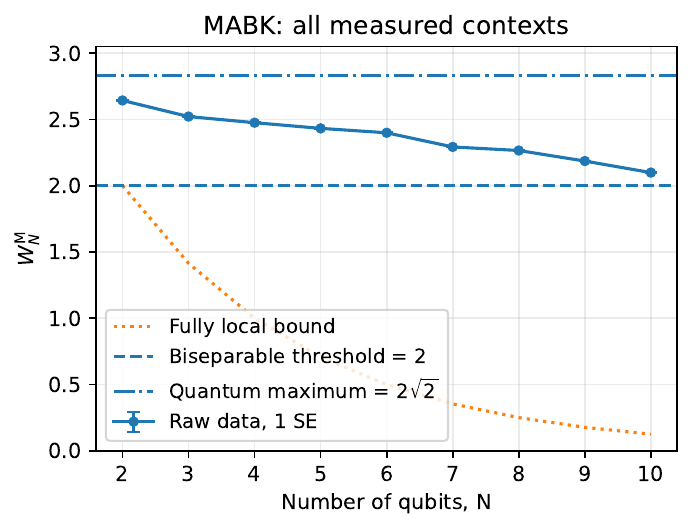}
\caption{MABK data and one-standard-error bars on the scale
\eqref{eq:ibm-scale}. The local bound decreases with $N$; the horizontal
line at $2$ is the biseparable threshold, not the fully local bound.
Error bars are smaller than the markers at this scale.}
\label{fig:ibm-mabk}
\end{figure}
\begin{figure}[tbp]
\includegraphics[width=\columnwidth]{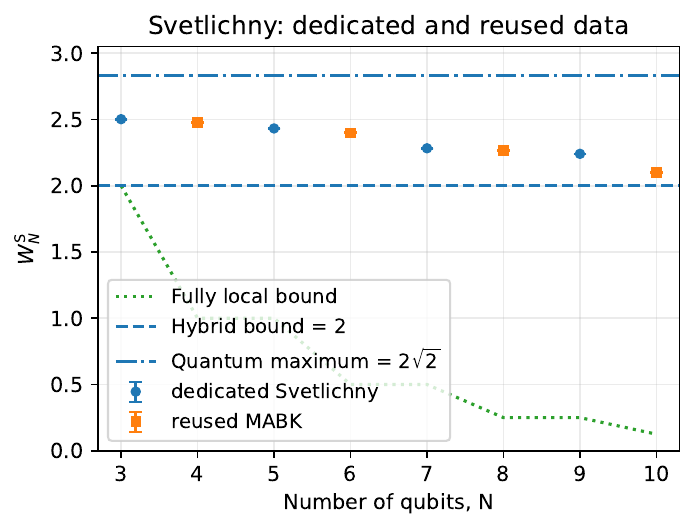}
\caption{Svetlichny scores with one-standard-error bars. Squares reuse
the even-$N$ MABK data after the fixed relabeling; circles are the separate
odd-$N$ jobs. The line at $2$ is the hybrid bound. Error bars are smaller
than the markers at this scale.}
\label{fig:ibm-svet}
\end{figure}
\begin{figure}[tbp]
\includegraphics[width=\columnwidth]{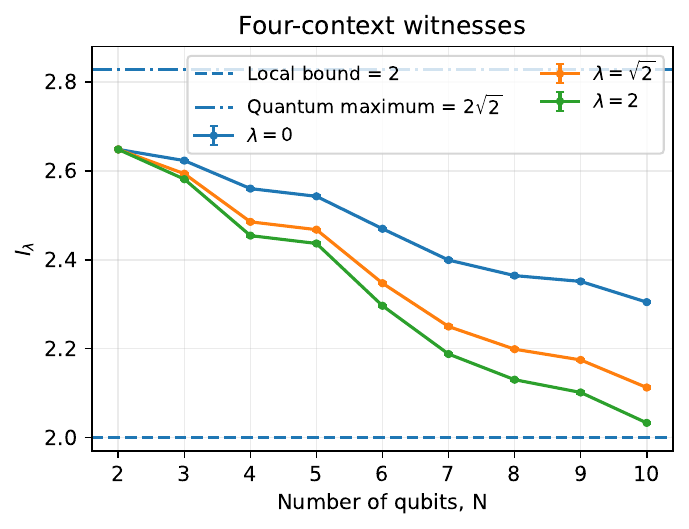}
\caption{Four-context estimates at $\lambda=0,\sqrt2,2$, computed from
the same counts at each $N$. The local bound is $2$ and the quantum maximum
is $2\sqrt2$. GME requires the penalty-dependent bound in
Fig.~\ref{fig:ibm-lambda}, not merely the local bound.}
\label{fig:ibm-four}
\end{figure}
\begin{figure}[tbp]
\includegraphics[width=\columnwidth]{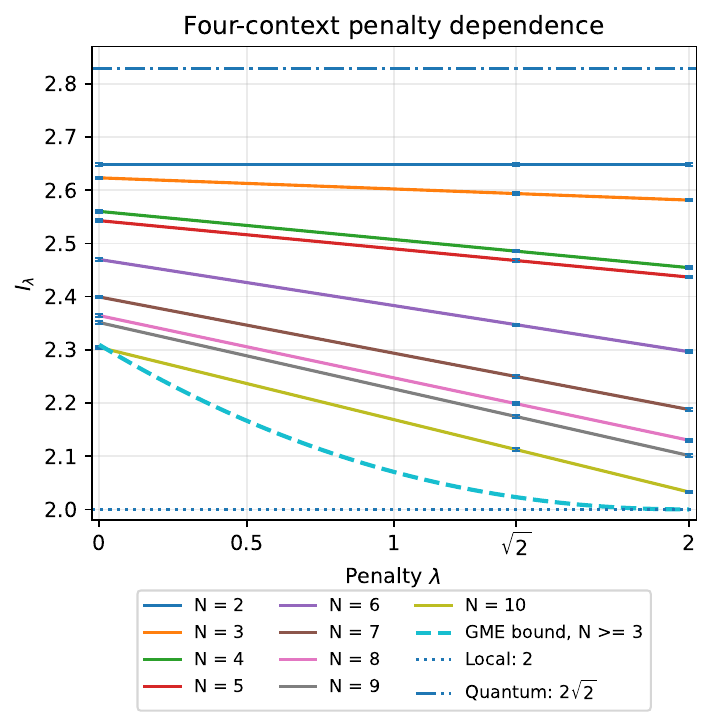}
\caption{Penalty dependence $I_\lambda=I_0-
\lambda(1-p)$ and the biseparable bound for $N\geq3$. Shading and
error bars indicate pointwise shot-noise standard errors, not a simultaneous
confidence band. Varying $\lambda$ requires no additional circuits. The
$N=2$ curve is constant because a single Bob is always unanimous.}
\label{fig:ibm-lambda}
\end{figure}

For all $3\leq N\leq10$, the measured MABK and Svetlichny means exceed
their corresponding threshold $2$. At ten qubits,
\begin{align}\label{eq:ibm-ten}
 W_{10}^{\rm M}= W_{10}^{\rm S}
 &=2.09865\pm0.00231,\notag\\
 I_2&=2.03276\pm0.00238.
\end{align}
The latter also exceeds the sharp bipartite-source network bound of
Theorem~\ref{thm:triangle}, under that model's stated assumptions. In contrast,
$I_0=2.30456\pm0.00214$ at $N=10$ lies below $4/\sqrt3$ and
therefore does not certify GME at that size. Both $I_{\sqrt2}$ and $I_2$
exceed their GME thresholds through $N=10$. For every measured
$3\leq N\leq10$, $I_2$ also exceeds the bipartite-source network bound.
These are comparisons with
specified mathematical null models, subject to the implementation and
statistical limitations below; a positive mean margin alone is not an
assumption-free confidence statement.

\subsection{Error estimation and limitations}
For each trial in a $y=0$ context, let $t=a\prod_jb_j$. In a $y=1$ context
let $h$ be the common Bob sign on unanimity and zero otherwise, and put
$c=h^2$, $t=(-1)^xah$. A directly observable score is
\begin{equation}\label{eq:shotscore}
 f_{xy}(\lambda)=
 \begin{cases}t,&y=0,\\t-\frac\lambda2(1-c),&y=1.\end{cases}
\end{equation}
The estimator is
\begin{equation}\label{eq:estimator}
\widehat I_\lambda=\sum_{x,y}\overline{f_{xy}(\lambda)}.
\end{equation}
The overbar denotes the corresponding sample mean. The estimator averages the two measured unanimity frequencies and incorporates their
covariance with the correlators. For complete setting sweeps with equal
shots per circuit, a context mean is the average of its circuit means.
Writing $\omega_u=n_u/n_s$ for the fraction of a setting's shots in circuit
$u$, the reported variance estimator is
\begin{equation}\label{eq:variance}
\widehat{\Var}(\widehat I_\lambda)
 =\sum_s\sum_{u\in s}\omega_u^2\frac{\widehat{\Var}_u(f_s)}{n_u}.
\end{equation}
Here $\widehat{\Var}_u(f_s)$ is the unbiased sample variance of the
individual-shot scores in circuit $u$. For a full-correlation polynomial, the analogous expression uses the
parity score, with squared Bell coefficients and the squared normalization
factor. This is a within-circuit shot-noise estimate, conditional on
independent shots; it is not a total hardware uncertainty.

We additionally report the standard error across complete submission
sweeps where repetitions exist. For MABK at $N=5$ and $7$, the observed
sweep-variance/shot-variance ratios are $2.63$ and $3.46$; for dedicated
Svetlichny at $N=5$ the ratio is $2.14$. These are descriptive drift or
heterogeneity diagnostics, particularly uncertain when only a few sweeps
are available. They warn against interpreting the tabulated shot errors
as a universal noise model. The one-sweep runs cannot support this check.
Single-party marginal consistency checks in the four-context runs have
maximum absolute standardized contrast $2.80$ over $108$ comparisons;
this limited diagnostic neither proves no-signaling nor closes a loophole.

No mitigation, readout inversion, postselection, or model-based count
correction enters the tables. Nonetheless, measurements on one processor
are not spatially separated, inputs are submitted in circuits rather than
chosen independently on every shot, and archived histograms do not retain
the chronological shot sequence. Thus these experiments benchmark
DI witness functionals under a local-measurement/common-preparation model;
they are not loophole-free or arbitrary-memory DI demonstrations
\cite{Zhang2011}.

\subsection{A common, explicitly stated noise benchmark}
For theory-only comparisons, take global white noise
\begin{equation}\label{eq:whitenoise}
 \rho_\nu=\nu\proj{\GHZ_N(\phi)}+(1-\nu)\frac{1}{2^N},
 \qquad0\leq\nu\leq1.
\end{equation}
This is a benchmark model, not a characterization of IBM hardware noise.
For the phase-matched axes \eqref{eq:fouraxes}, put $q_N=2^{2-N}$. Then
\begin{equation}\label{eq:noiseI}
 p=\nu+(1-\nu)q_N,\qquad
 I_\lambda=2\sqrt2\,\nu-\lambda(1-\nu)(1-q_N).
\end{equation}
For $N\geq3$, the GME visibility threshold is
\begin{equation}\label{eq:noisecritical}
 \nu>\frac{b_\BS(\lambda)+\lambda(1-q_N)}
 {2\sqrt2+\lambda(1-q_N)}.
\end{equation}
Thus an $N$-independent score threshold is not an $N$-independent tolerance
to arbitrary noise. In the same global-white-noise model, MABK GME and
Svetlichny GMNL both require $\nu>1/\sqrt2$, while
$\mathcal K_N$ requires $\nu>m/\sqrt{1+m^2}$.
For example, $I_0$ has threshold $\sqrt{2/3}$, and the $I_2$ threshold tends
to $2(\sqrt2-1)$ as $N\to\infty$.

\section{Discussion}\label{sec:discussion}
Four binary-outcome contexts suffice for GHZ GME, but not for
Svetlichny-GMNL. The sparse GHZ test saves contexts at the cost of a smaller
violation margin; for even $N$, MABK and Svetlichny are equivalent after
relabeling. Bell-tree GME also needs only four contexts, but with one output
bit per incident edge. Its chained GMNL certificate requires a different
input schedule to prevent setting leakage within hybrid groups.

These results separate multipartite entanglement and nonlocality from the
question of bipartite-source preparation. The exact triangle proof and
monotone merging establish the sharp network bound $I_\lambda\leq2$ for
every $N\geq3$ and $\lambda\geq2$, including shared-randomness mixtures.
Thus $I_2>2$ both certifies GME and excludes this network class, whereas
Bell-pair trees demonstrate that GME and GMNL can also occur inside the
class. The selected IBM data
provide raw-count witness benchmarks through ten qubits. Their interpretation
retains the distinction between shot-noise uncertainty, hardware drift,
and the operational assumptions of device-independent certification.

\section*{Acknowledgements}
The results have been created using IBM Quantum. The views expressed are those of the
authors and do not reflect the official policy or position of the IBM Quantum. We also acknowledge the use of ChatGPT 6  (OpenAI, https://chat.openai.com) for assistance with grammar polishing and for text-based verification during the development of some mathematical proofs. All AI-generated suggestions, logical transitions, and derivations were manually cross-checked, independently re-verified, and approved by the human authors, who retain full accountability for the accuracy and integrity of the work. We thank Tomasz  Rybotycki and Piotr Gawron
for support and  insightful discussions.

\appendix
\section{Identity, Pauli operators, state phases, and dilation}\label{app:notation}
The identity is denoted by $1$, as specified in Sec.~\ref{sec:models}.
An empty tensor product or product of commuting local operators is also $1$.
The standard Pauli matrices in the ordered basis $\ket0,\ket1$ are
\begin{equation}\label{eq:pauli}
 X=\begin{pmatrix}0&1\\1&0\end{pmatrix},\quad
 Y=\begin{pmatrix}0&-\ii\\\ii&0\end{pmatrix},\quad
 Z=\begin{pmatrix}1&0\\0&-1\end{pmatrix}.
\end{equation}
They satisfy $X^2=Y^2=Z^2=1$, $XY=\ii Z$, and cyclic permutations.
We use $[F,G]=FG-GF$ and $\{F,G\}=FG+GF$.
For an equatorial angle $\alpha$,
\begin{equation}\label{eq:Xalpha}
 \tilde{X}_\alpha=X\cos\alpha+Y\sin\alpha.
\end{equation}
The single phase-dependent definition used throughout is
\begin{equation}\label{eq:ghzdef}
 \ket{\GHZ_N(\phi)}=
 \frac{\ket{0^N}+e^{\ii\phi}\ket{1^N}}{\sqrt2},\qquad
 \ket{r^N}=\bigotimes_{j=1}^N\ket r.
\end{equation}
The Bell-family notation is an abbreviation of this same definition:
\begin{equation}\label{eq:belldef}
 \ket{\Phi(\phi)}:=\ket{\GHZ_2(\phi)},\qquad
 \ket{\Phi^+}:=\ket{\Phi(0)}.
\end{equation}
In particular,
\begin{align}\label{eq:equatorial}
 \avg{\bigotimes_{j=1}^N \tilde{X}_{\alpha_j}}_{\GHZ_N(\phi)}
 &=\cos\!\left(\phi-\sum_j\alpha_j\right),\nonumber\\
 \avg{\bigotimes_j(X+\ii Y)}_{\GHZ_N(\phi)}
 &=2^{N-1}e^{\ii\phi}.
\end{align}
The signs of all phase-matching prescriptions in the main text follow from
these equations.

For any Hermitian contraction $O_x$, the operator on a doubled local space
\begin{equation}\label{eq:dilation}
 \widehat O_x=\begin{pmatrix}
 O_x&\sqrt{1-O_x^2}\\
 \sqrt{1-O_x^2}&-O_x
 \end{pmatrix}
\end{equation}
is a reflection. The same doubled space and embedding
$V\ket\psi=\ket\psi\otimes\ket0$ work for every setting $x$:
$V^\dagger\widehat O_xV=O_x$ and
$V^\dagger(1\pm\widehat O_x)V/2=(1\pm O_x)/2$.
Taking independent local ancillas preserves all measurement probabilities,
all interlaboratory correlators, separability across any cut, and the
bipartite-source structure. No selected outcome is involved. Consequently,
proofs may use reflections without assuming projective hardware measurements.
Unmeasured same-party products, such as a squared difference of two
observables, refer to this dilation when it is used.

\section{Standard Bell bounds}\label{app:standard}
\subsection{Uffink bound and the MABK normalization}
Here we give a proof of \eqref{eq:uffink}, following the quadratic-bound
approach of Ref.~\cite{uff}, with arbitrary dimensions covered by local
dilation and the two-reflection decomposition.  
On a nontrivial two-dimensional invariant block, choose coordinates such that
\begin{equation}
 A_0=X\cos\theta+Y\sin\theta,\qquad
 A_1=X\cos\theta-Y\sin\theta.
\end{equation}
Then
\begin{equation}
 A_0+\ii A_1=2e^{\ii\pi/4}
 \begin{pmatrix}0&u\\v&0\end{pmatrix},\qquad
 u^2+v^2=1,
\end{equation}
where $u=(\cos\theta-\sin\theta)/\sqrt2$ and
$v=(\cos\theta+\sin\theta)/\sqrt2$ are real. Commuting one-dimensional
blocks can be embedded in the same description.

For a tensor product of $N$ such blocks, remove the factor
$2^Ne^{\ii N\pi/4}$. The remaining operator splits into two-dimensional
blocks pairing each bitstring with its complement. Its two off-diagonal
entries are $a=\prod_jc_j$ and $b=\prod_jd_j$, where
$c_j^2+d_j^2=1$. On a normalized vector, the modulus of the expectation
of that block is at most $(|a|+|b|)/2$. For $N\geq2$,
\begin{align}
 (|a|+|b|)^2
 &\leq(c_1^2+d_1^2)
 \left(\prod_{j>1}c_j^2+\prod_{j>1}d_j^2\right)\nonumber\\
 &\leq c_2^2+d_2^2=1.
\end{align}
Averaging over blocks and states proves
$|\avg{\mathcal M_N}|\leq2^{N-1}$.
A single party instead obeys
$|\avg{A_0}+\ii\avg{A_1}|\leq\sqrt2$.
For infinite-dimensional trace-class states, local finite-rank compression,
dilation of the compressed contractions, and continuity extend the bound.

For deterministic local outputs, each factor $a_0+\ii a_1$ has modulus
$\sqrt2$ and phase $\pi/4$ modulo $\pi/2$. After multiplication and the
phase shift in \eqref{eq:mabk}, the phase is again an odd multiple of
$\pi/4$. Hence the maximal real part is $2^{(N-1)/2}$ before normalization,
and the local bound is $1$. The same coefficient phase shows why the
number of nonzero terms is $2^{N-1}$ for odd $N$ and $2^N$ for even $N$.

For a product partition with $s$ singleton blocks and $g$ larger blocks,
the complex expectation is bounded by
$2^{s/2}\prod_{\ell=1}^g2^{n_\ell-1}=2^{N-s/2-g}$.
Multiplying by $2^{(1-N)/2}$ gives \eqref{eq:partitionbound}.
For a bipartition, a singleton versus an $(N-1)$-party block gives
$2^{(N-2)/2}$, while two blocks of size at least two give
$2^{(N-3)/2}$. GHZ phases in the larger block can align the complex
expectation, so the former is tight for $N\geq3$.
Convexity includes mixtures over cuts. Fully separable states use the
sharper deterministic-local argument above.

\subsection{Local and quantum bounds of the four-context family}\label{app:fourbasic}
For deterministic outputs, let $r=\prod_jb_{j,0}$. On unanimous input-$1$
outcomes, let $h$ be their common sign. Then
$I_\lambda=(a_0+a_1)r+(a_0-a_1)h\leq2$.
On disagreement, $I_\lambda=(a_0+a_1)r-\lambda\leq2$ for $\lambda\geq0$.
All $+1$ outputs attain $2$.

After \eqref{eq:dilation}, $R_0^2=1$, $P$ is a projection,
$\Gamma^2=P$, and $(A^+)^2+(A^-)^2=4$.
For a pure state or a purification, Cauchy--Schwarz applied to
$(A^+\ket\psi,A^-\ket\psi)$ and
$(R_0\ket\psi,\Gamma\ket\psi)$ gives
\begin{equation}\label{eq:quantumtradeoff}
 \mathcal T\leq2\sqrt{1+p},\qquad
 I_\lambda\leq2\sqrt{1+p}-\lambda(1-p)\leq2\sqrt2.
\end{equation}
The penalty is nonpositive, since $1-p\geq0$.
Equation~\eqref{eq:fourghz} establishes attainment. This is consistent with
the CHSH quantum bound \cite{chsh,tsirel}; treating the Bobs as one
unrestricted laboratory by itself would prove only a bipartite statement,
not the biseparable bound.

\subsection{CHSH marginal-bias tradeoff}\label{app:bias}
We prove the second estimate in \eqref{eq:Tbounds}, a form of the standard
CHSH bias tradeoff used in Ref.~\cite{pironio}.
In a two-dimensional Alice block, write
$A^+=2X\cos\theta$, $A^-=2Z\sin\theta$ and define
\begin{equation}
 r=\avg{XR_0},\quad z=\avg Z,\quad
 v=\avg{Z\Gamma},\quad u=\avg\Gamma.
\end{equation}
Cauchy--Schwarz gives $\mathcal T^2/4\leq r^2+v^2$.
The reflections $XR_0$ and $Z$ anticommute, so $r^2+z^2\leq1$.
Since $\Gamma$ is a contraction commuting with $Z$,
\begin{equation}
 |u+v|\leq1+z,\qquad |u-v|\leq1-z.
\end{equation}
Squaring and adding yields $u^2+v^2\leq1+z^2$, hence
$\mathcal T^2/4+u^2\leq2$.
One-dimensional blocks have $|\mathcal T|\leq2$, $|u|\leq1$ and satisfy the same
inequality. The disk $\mathcal T^2/4+u^2\leq2$ is convex, so averaging over Alice
blocks proves $\mathcal T\leq2\sqrt{2-u^2}$ in the original state.
The block states are used only for this universal quantum bound; no Finner
inequality is applied to a postselected Jordan block.

\subsection{Svetlichny bounds}\label{app:svet}
For deterministic local outcomes, the phase argument above, now applied to
$(1+\ii)\prod_j(a_{j,0}+\ii a_{j,1})$, gives local maximum
$2^{\lfloor(N+1)/2\rfloor}$. For the hybrid bound, fix a cut with $\ell$ and
$N-\ell$ parties. Let $F_r,F_i$ be the signed parity sums of the even- and
odd-weight inputs on the first side, and $G_r,G_i$ the analogous sums on
the second. Within a group its parity may be an arbitrary function of all
its inputs. Nevertheless,
\begin{equation}
 |F_r|,|F_i|\leq2^{\ell-1},\qquad
 |G_r|,|G_i|\leq2^{N-\ell-1}.
\end{equation}
The score factorizes as
\begin{equation}
 \mathcal S_N=F_r(G_r-G_i)-F_i(G_r+G_i).
\end{equation}
Maximizing this bilinear expression over the corresponding box gives
$2\cdot2^{\ell-1}\cdot2^{N-\ell-1}=2^{N-1}$.
The bound is achievable by group parity assignments and extends to mixtures
of cuts. This is the Svetlichny argument \cite{svetlichny,seevinck,collinsm4}.

For quantum states, \eqref{eq:uffink} implies
$|\mathcal S_N|\leq\sqrt2\,2^{N-1}$, attained as stated in the main text.
For quantum-biseparable states and $N\geq3$, the product of the singleton
bound $\sqrt2$ and the $(N-1)$-party Uffink bound, with the prefactor
$\sqrt2$, yields $2^{N-1}$; two nonsingleton blocks give a smaller bound.
A deterministic singleton and a suitably phased GHZ state attain the maximum.

For even $N$, $\vartheta_N$ is an odd multiple of $\pi/4$; hence the
nonzero MABK coefficients all have modulus $2^{-N/2}$ and match a
Svetlichny sign pattern up to output relabelings. For odd $N$, MABK
selects one quadrature of $\avg{\mathcal M_N}$ instead. The group parity
box gives maximum $2^{N-1}$ for either quadrature, yielding the normalized
hybrid bound $2^{(N-1)/2}$. Both box extrema are attainable because the
groups' even- and odd-input parity tables are independently assignable.

\subsection{Chained inequality: local and quantum bounds}\label{app:chain}
Suppress the edge index in \eqref{eq:chain}. The $2L$ signed terms cannot
all equal $+1$ under a local deterministic assignment: multiplying the
required signs would give $1=-1$. Thus $S_L\leq2L-2$, attained with all
outputs $+1$. Its algebraic bound is $2L$.

For the standard quantum bound \cite{wehner}, take reflections on a pure
state and realify the Hilbert-space inner product. The unit vectors
$u_i=A_i\ket\psi$, $v_i=B_i\ket\psi$ satisfy
$u_i\cdot v_j=\avg{A_iB_j}$. Put $u_{L+1}=-u_1$. Then
\begin{equation}
 |S_L|=\left|\sum_{i=1}^L(u_i+u_{i+1})\cdot v_i\right|
 \leq\sum_{i=1}^L\norm{u_i+u_{i+1}}.
\end{equation}
Writing $\theta_i=\arccos(u_i\cdot u_{i+1})$, angular distance obeys
$\sum_i\theta_i\geq\pi$ because the chain ends at the antipode of $u_1$.
Concavity and monotonicity of $\cos(\theta/2)$ on $[0,\pi]$ give
\begin{equation}
 |S_L|\leq2\sum_i\cos(\theta_i/2)
 \leq2L\cos\!\left(\frac{\sum_i\theta_i}{2L}\right)
 \leq2L\cos\frac\pi{2L}.
\end{equation}
Dilation and convexity cover arbitrary binary POVMs and mixed states.
On $\ket{\Phi^+}$, the bound is attained by
\begin{align}\label{eq:chainaxes}
 A_i&=Z\cos\alpha_i+X\sin\alpha_i,&\alpha_i&=\frac{(i-1)\pi}{L},\nonumber\\
 B_i&=Z\cos\beta_i+X\sin\beta_i,&\beta_i&=\frac{(2i-1)\pi}{2L}.
\end{align}
Here $\avg{A_iB_j}=\cos(\alpha_i-\beta_j)$, so every signed term equals
$\cos(\pi/2L)$. A tensor product of these measurements on incident qubits
attains the tree quantum value without requiring joint measurements
between laboratories.

\section{Complete tight biseparable optimization}\label{app:bisep}
\subsection{Reduction of an arbitrary cut}
It is enough to treat a product state across
$L=\{A\}\cup S$ and $R=\mathsf J\setminus S\neq\varnothing$.
The set $S$ can be empty. After local dilation, let
\begin{equation}
 R_{0S}=\prod_{j\in S}B^{(j)}_0,\qquad
 P_{S\pm}=\prod_{j\in S}B^{(j)}_{\pm|1},
\end{equation}
and use the analogous notation for $R$. Set
$u_R=\avg{R_{0R}}$, $r_\pm=\avg{P_{R\pm}}$.
Because $R$ is nonempty,
$|u_R|\leq1$, $r_\pm\geq0$, and $r_++r_-\leq1$.
Factorization across the cut gives
\begin{align}\label{eq:cutreduction}
 I_\lambda={}&u_R\avg{A^+R_{0S}}
 +r_+\avg{(A^-+\lambda)P_{S+}}\nonumber\\
 &+r_-\avg{(-A^-+\lambda)P_{S-}}-\lambda.
\end{align}
Relax the attainable triples to the interval times the full simplex.
Linearity reduces the maximum to
$u_R=\pm1$ and $(r_+,r_-)\in\{(0,0),(1,0),(0,1)\}$.
The zero vertex gives at most $2-\lambda$.
At the plus vertex absorb $u_R$ into a contraction $V$, put $Q=P_{S+}$,
and relax $V,Q$ to arbitrary operators on the joint $S$ space. The problem is
\begin{equation}\label{eq:Klambda}
 K_\lambda=A^+V+A^-Q-\lambda(1-Q),
\end{equation}
where $V=V^\dagger$, $\norm V\leq1$, and $Q^2=Q$.
The minus vertex has the same bound after $A^-\mapsto-A^-$.
Scalar $V,Q$ include $S=\varnothing$.
For $\lambda=2$, the short operator argument following
\eqref{eq:shortK} already proves $K_2\preceq2$.

\subsection{Uniform bound for \texorpdfstring{$0\leq\lambda\leq2$}{0 <= lambda <= 2}}
The two-reflection decomposition of Alice's observables gives, on each
nontrivial block,
\begin{equation}\label{eq:AliceJordan}
 A^+=2\sqrt{1-s^2}\,X,\qquad A^-=2sZ,\qquad0\leq s\leq1.
\end{equation}
One-dimensional blocks contribute at most $2$ directly.
For a proposed upper bound $b\geq2$, the diagonal blocks of $b-K_\lambda$
in the $Z$ basis are
\begin{align}
 L_0&=(b+\lambda)-(2s+\lambda)Q\succeq b-2s,\nonumber\\
 L_1&=(b+\lambda)+(2s-\lambda)Q
 \succeq b+\min\{\lambda,2s\}.
\end{align}
The off-diagonal block is $-2\sqrt{1-s^2}\,V$. Its quadratic form on
$(\xi,\zeta)$ is bounded below by
\begin{align}
 &(b-2s)\norm\xi^2+(b+\min\{\lambda,2s\})\norm\zeta^2\nonumber\\
 &\hspace{18mm}-4\sqrt{1-s^2}\norm\xi\norm\zeta.
\end{align}
Therefore $b-K_\lambda\succeq0$ if
\begin{equation}\label{eq:detcondition}
 (b-2s)(b+\min\{\lambda,2s\})\geq4(1-s^2).
\end{equation}
When $2s\leq\lambda$, this is implied by $b\geq2$.
When $2s\geq\lambda$, it suffices that
\begin{equation}\label{eq:flambda}
 b\geq f_\lambda(s):=s-\frac\lambda2+
 \sqrt{4-3s^2+\lambda s+\frac{\lambda^2}{4}}.
\end{equation}
For $0\leq\lambda<2$, this strictly concave function has its unique maximum
on $[\lambda/2,1]$ at
\begin{equation}\label{eq:sstar}
 s_* =\frac{\lambda+\sqrt{\lambda^2+12}}6,
 \qquad3s_*^2-\lambda s_*-1=0.
\end{equation}
Indeed, its stationary condition is
$2\sqrt{4-3s^2+\lambda s+\lambda^2/4}=6s-\lambda$.
The solution lies in the required interval for $0\leq\lambda\leq2$.
At this point,
\begin{equation}
 f_\lambda(s_*)=4s_*-\lambda
 =\frac{2\sqrt{\lambda^2+12}-\lambda}{3}\geq2.
\end{equation}
The endpoint $\lambda=2$ follows directly or by continuity.
Thus $b=b_\BS(\lambda)$ bounds all blocks, all vertices in
\eqref{eq:cutreduction}, and every cut. Convexity then covers arbitrary
biseparable mixtures. Local finite-rank compression and continuity extend
the finite-dimensional argument to arbitrary trace-class realizations.

\subsection{Attaining biseparable strategy and boundary cases}
On $AB^{(2)}$ use
\begin{equation}\label{eq:bsstate}
 \ket{\psi_*}=\sqrt{w_*}\ket{00}+\sqrt{1-w_*}\ket{11},
 \qquad w_* =\frac{4s_*^2}{1+3s_*^2},
\end{equation}
and put every remaining qubit in $\ket0$. Choose
\begin{equation}
 A_x=\sqrt{1-s_*^2}\,X+(-1)^xs_*Z,
\end{equation}
$B^{(2)}_0=X$, $B^{(2)}_1=Z$, and $Z$ for both settings of every spectator.
In the $\ket{00},\ket{11}$ subspace the score operator is
\begin{equation}\label{eq:bsmatrix}
 \begin{pmatrix}
 2s_*&2\sqrt{1-s_*^2}\\
 2\sqrt{1-s_*^2}&-\lambda
 \end{pmatrix}.
\end{equation}
Its top eigenvalue is $f_\lambda(s_*)$ and its eigenvector amplitude ratio
is $\sqrt{1-s_*^2}/(2s_*)$, which gives \eqref{eq:bsstate}.
Hence the upper bound is tight for every $N\geq3$.
For $\lambda\geq2$, $I_\lambda\leq I_2\leq2$ on biseparable states,
and a deterministic strategy attains $2$.
No $\lambda>2$ improves the violation margin over $\lambda=2$ on the same
data. For two parties, the unanimity event is certain and the family is
exactly CHSH.

\section{A single Finner and tensor-contraction toolbox}\label{app:finner}
All Finner applications in the paper refer to this appendix. The classical
inequality is due to Finner \cite{Finner1992}; the quantum positive-operator
version for networks with bipartite sources is given by Renou
\emph{et al.}~\cite{Renou2019}. The explicit non-Hermitian step below is
included because the MABK application requires it.

\subsection{Closed graph contraction}
\begin{lemma}\label{lem:graph}
For a finite loopless graph, let each edge $e$ carry a finite index $s_e$,
and let the complex array $T_v$ depend on the indices incident to $v$.
Then
\begin{equation}\label{eq:graphnorm}
 \left|\sum_{(s_e)}\prod_vT_v((s_e)_{e\ni v})\right|
 \leq\prod_v\norm{T_v}_F,
\end{equation}
where $\norm{T_v}_F^2$ sums the squared absolute values of all its entries.
\end{lemma}
\begin{proof}
Pad missing edges with one-element index sets; parallel edges can be
combined into a composite index. It suffices to prove the complete-graph
case by induction. For two vertices the claim is Cauchy--Schwarz.
For $n$ vertices, let $x_j=s_{1j}$, $j\geq2$, and let
$\boldsymbol y$ collect all remaining edge indices. For fixed
$\boldsymbol y$, Cauchy--Schwarz over $\boldsymbol x$ gives
\begin{align}
 &\left|\sum_{\boldsymbol x}T_1(\boldsymbol x)
 \prod_{j=2}^nT_j(x_j,\boldsymbol y_j)\right|\nonumber\\
 &\quad\leq\norm{T_1}_F
 \prod_{j=2}^n\left(\sum_{x_j}|T_j(x_j,\boldsymbol y_j)|^2\right)^{1/2}.
\end{align}
Each $x_j$ occurs in only one remaining factor, so the sum factorizes.
Define the square-root factors as reduced nonnegative arrays
$\widehat T_j(\boldsymbol y_j)$. Their Frobenius norms equal
$\norm{T_j}_F$. Summing over $\boldsymbol y$ and applying the induction
hypothesis proves the result. The one-vertex case is a scalar equality.
\end{proof}
Equivalently, \eqref{eq:graphnorm} is the counting-measure Finner inequality
with vertex weights $1/2$, since each edge variable occurs at two vertices.
The reduced contraction is a bound, not an equality with the original one.

\subsection{Quantum operator contraction}
Let each source be pure, in Schmidt form
\begin{equation}
 \ket{\psi_e}=\sum_i\sqrt{p_{e,i}}\ket{i}_{u|v}\ket{i}_{v|u},
 \qquad\sum_i p_{e,i}=1,
\end{equation}
and put $\ket\Psi=\bigotimes_e\ket{\psi_e}$.
Let $\sigma_v$ be the actual local reduced state of this pure network.
\begin{lemma}\label{lem:sandwich}
For arbitrary bounded local operators $F_v$, not necessarily Hermitian,
\begin{equation}\label{eq:sandwich}
 \left|\bra\Psi\bigotimes_vF_v\ket\Psi\right|
 \leq\prod_v\norm{\sigma_v^{1/4}F_v\sigma_v^{1/4}}_2.
\end{equation}
Here $\norm K_2^2=\Tr(K^\dagger K)$ is the Hilbert--Schmidt norm.
\end{lemma}
\begin{proof}
For edge $e$, use a doubled index $(i_e,j_e)$ for the bra and ket Schmidt
labels. Define
\begin{equation}
 T_v((i_e,j_e)_{e\ni v})=(F_v)_{\boldsymbol i,\boldsymbol j}
 \prod_{e\ni v}(p_{e,i_e}p_{e,j_e})^{1/4}.
\end{equation}
The expectation is exactly the closed contraction $\sum\prod_vT_v$.
In the product Schmidt basis,
$(\sigma_v)_{\boldsymbol i,\boldsymbol i}=\prod_{e\ni v}p_{e,i_e}$,
so $\norm{T_v}_F=\norm{\sigma_v^{1/4}F_v\sigma_v^{1/4}}_2$.
Lemma~\ref{lem:graph} proves the result without dimension factors.
\end{proof}
For Hermitian $H$,
\begin{align}\label{eq:WY}
 \Tr(\sigma H^2)-\Tr(\sqrt\sigma H\sqrt\sigma H)
 &=\tfrac12\norm{[\sqrt\sigma,H]}_2^2\geq0.
\end{align}
This is the Wigner--Yanase skew-information identity
\cite{WignerYanase1963}. In particular, for contractions $A_0,A_1$,
\begin{align}\label{eq:localbudget}
 &\norm{\sigma^{1/4}(A_0+\ii A_1)\sigma^{1/4}}_2^2\nonumber\\
 &\quad=\sum_{x=0}^1\Tr(\sqrt\sigma A_x\sqrt\sigma A_x)
 \leq\sum_{x=0}^1\Tr(\sigma A_x^2)\leq2.
\end{align}
The cross term is the trace of a commutator and vanishes; anticommutation
of the observables is not assumed.
Substitution in \eqref{eq:sandwich} proves
$|\avg{\mathcal M_N}|\leq2^{N/2}$ for pure sources.
Mixed independent sources can be purified by assigning each purifying
system to one endpoint and extending local measurements by identity.
The same local budget applies on those enlarged spaces. Alternatively,
pure-source decompositions and convexity give the same bound.
Finally,
\begin{equation}
 \left|\sum_\eta p_\eta\avg{\mathcal M_N}_\eta\right|
 \leq\sum_\eta p_\eta2^{N/2}=2^{N/2}.
\end{equation}
This proves Theorem~\ref{thm:networkM}, including shared randomness.

\subsection{Positive effects and functional Finner}
\begin{lemma}[Quantum Finner, weights $1/2$]\label{lem:finner}
For a fixed independent bipartite-source network and bounded positive
local operators $F_v$,
\begin{equation}\label{eq:functionalfinner}
 \avg{\bigotimes_vF_v}\leq\prod_v\avg{F_v^2}^{1/2}.
\end{equation}
For effects $0\preceq E_v\preceq1$, this implies
$\avg{\bigotimes_vE_v}\leq\prod_v\avg{E_v}^{1/2}$.
\end{lemma}
\begin{proof}
Apply \eqref{eq:sandwich}, followed by \eqref{eq:WY} to each Hermitian $F_v$.
For effects use $E_v^2\preceq E_v$. Purifying independent mixed sources at
one endpoint per edge preserves both sides of \eqref{eq:functionalfinner}.
\end{proof}
Local positive operators can also be obtained by spectral measurement and
nonnegative postprocessing, giving the functional form of Theorem~2 of
Ref.~\cite{Renou2019}. Local joint measurements on several incoming
systems are included. All these finite-dimensional arguments extend to
bounded measurements on trace-class states by finite-rank Schmidt
truncation and continuity.

Shared-randomness averaging need not preserve \eqref{eq:functionalfinner}.
For example, three parties outputting a common unbiased shared bit have
$P(000)=1/2>(1/2)^{3/2}$. This is why the main-text network witnesses first
obtain a constant score bound inside each component.

\subsection{A covariance consequence for independent triangles}\label{app:covariance}
\begin{lemma}\label{lem:covariance}
For bounded local Hermitian operators $M_A,M_B,M_C$ in an independent-source
triangle, the matrix with diagonal entries $\Var(M_j)$ and off-diagonal
entries $-\Cov(M_i,M_j)$ is positive semidefinite.
\end{lemma}
\begin{proof}
Let $\Delta_j=M_j-\avg{M_j}$, and choose any real weights $t_j$.
The local operators $F_j(\varepsilon)=1+\varepsilon t_j\Delta_j$ are
positive for sufficiently small $|\varepsilon|$.
The left side of \eqref{eq:functionalfinner} is
\begin{equation}
 1+\varepsilon^2\sum_{i<j}t_it_j\Cov(M_i,M_j)+O(\varepsilon^3),
\end{equation}
whereas the right side is
\begin{equation}
 1+\frac{\varepsilon^2}{2}\sum_jt_j^2\Var(M_j)+O(\varepsilon^4).
\end{equation}
Subtract $1$, divide by $\varepsilon^2$, and take $\varepsilon\to0$.
The resulting quadratic form is nonnegative for every real vector
$(t_A,t_B,t_C)$, proving the claim.
\end{proof}
The minus signs on the off-diagonal covariances distinguish this network
constraint from ordinary covariance-matrix positivity. It applies to local
Hermitian combinations such as $(A_0-A_1)/2$ without asserting that $A_0$
and $A_1$ can be measured jointly. This is a separate network
constraint; the exact proof below does not require it.

\section{Exact triangle bound}\label{app:triangle}

Fix one independent-source component of the model in
Sec.~\ref{sec:models}, and write $B=B^{(2)}$, $C=B^{(3)}$ for the two
nonpivot laboratories. The source state is
\begin{equation}\label{eq:sources}
 \rho=\rho_{AB}\otimes\rho_{AC}\otimes\rho_{BC}.
\end{equation}
Each party may measure its two incoming systems jointly. We retain the
$\pm1$ outcome convention of \eqref{eq:effects}: a binary observable $O$
is a Hermitian contraction, with effects $E_\pm=(1\pm O)/2$.
For a reflection, $O^2=1$, these effects are projections.

For three parties, \eqref{eq:aggregate} gives
\begin{equation}\label{eq:triangle-aggregates}
 R_0=B_0C_0,\qquad
 \Gamma=\frac{B_1+C_1}{2},\qquad
 P=\frac{1+B_1C_1}{2}.
\end{equation}
These identities hold for arbitrary binary effects. Substitution into
\eqref{eq:Ilambda} at $\lambda=2$ gives the score below.

\begin{theorem}[Exact triangle bound]\label{thm:triangle-bias}
For the independent sources \eqref{eq:sources} and arbitrary local binary
observables $A_0,A_1,B_0,B_1,C_0,C_1$,
\begin{equation}
\begin{aligned}
 I_2={}&\avg{(A_0+A_1)B_0C_0}\\
 &+\frac12\avg{(A_0-A_1)(B_1+C_1)}\\
 &+\avg{B_1C_1}-1\ \leq\ 2.
\end{aligned}\label{eq:triangle-score}
\end{equation}
The result holds in finite dimensions and for normal trace-class source
states on arbitrary Hilbert spaces. For every fixed bias
$\mu=(\avg{A_0}-\avg{A_1})/2\in[-1,1]$, the maximum is exactly $2$, attained
by a single independent-source realization.
\end{theorem}

We first prove the finite-dimensional statement with Alice's observables
Hermitian involutions, $A_0^2=A_1^2=\id$. Bob's and Charlie's observables
remain arbitrary Hermitian contractions. Appendix~\ref{app:triangle-extensions}
removes the restriction on Alice and gives the infinite-dimensional
extension. Set
\begin{equation}
 \begin{gathered}
 C_A=\frac{A_0+A_1}{2},\qquad D_A=\frac{A_0-A_1}{2},\\
 C_A^2+D_A^2=\id,\qquad C_A D_A+D_A C_A=0.
 \end{gathered}
 \label{eq:triangle-pivot}
\end{equation}
The last two identities hold under the stated involution assumption on
Alice. These definitions agree with $C_A=A^+/2$ and $D_A=A^-/2$
in the main-text notation, and $\mu=\avg{D_A}$.

All norms below are ordinary Hilbert-space or Euclidean vector norms.
The vector inner product $\inner{u}{v}$ is antilinear in its
first argument; $\avg{O}$ continues to denote a state expectation. Expectations may
be taken in mixed states. For every operator $R$,
\begin{equation}
 \begin{aligned}
 &\avg{R^\dagger R}-|\avg R|^2\\
 &\quad=\avg{(R-\avg R \id)^\dagger(R-\avg R \id)}\geq0.
 \end{aligned}
 \label{eq:column-variance}
\end{equation}
We use Cauchy--Schwarz for vectors and for the positive semidefinite form
$\avg{R^\dagger Q}$ induced by a state. The
particular positive identities used in the proof are given explicitly.

\subsection{The bipartite column inequality}
\label{app:triangle-column}

\begin{proposition}[Bipartite column inequality]\label{prop:column}
Let $\omega$ be any finite-dimensional bipartite state. On its first
factor let $A_0,A_1$ be Hermitian involutions, with $C_A,D_A$ as in
\eqref{eq:triangle-pivot}. On its second factor let $U,B$ be Hermitian contractions
and let $\mathsf R_1,\ldots,\mathsf R_{n_{\mathrm{col}}}$ be arbitrary
operators, where $n_{\mathrm{col}}$ is a nonnegative integer, satisfying
\begin{equation}
 D_{\rm col}:=\id-B^2-\sum_{i=1}^{n_{\mathrm{col}}} \mathsf R_i^\dagger \mathsf R_i\succeq0.
 \label{eq:column}
\end{equation}
Define real numbers $x,b$ and a complex vector $r$ by
\begin{equation}
 \begin{gathered}
 x=\avg{C_A\otimes U+D_A\otimes B},\\
 b=\avg{\id\otimes B},\qquad r_i=\avg{\id\otimes \mathsf R_i}.
 \end{gathered}
\end{equation}
Then
\begin{equation}
x^2+b^2+\norm r^2\leq2.
 \label{eq:column-bound}
\end{equation}
The case $n_{\mathrm{col}}=0$ is included, with an empty vector of norm zero.
\end{proposition}

\begin{proof}
\textbf{1. Spectral subspaces of $D_A^2$.}
The relation $C_A D_A=-D_A C_A$ implies $[C_A,D_A^2]=0$, and $[D_A,D_A^2]=0$ as well.
Decompose Alice's space into the spectral subspaces of $D_A^2$, using
the finite-dimensional spectral theorem.
On a subspace with eigenvalue $t^2$, where $0\leq t\leq1$, put
$s=\sqrt{1-t^2}$. The restricted operators satisfy
$D_A^2=t^2\id$ and $C_A^2=s^2\id$.

First suppose $s,t>0$. Set
\begin{equation}
 \mathsf Z=C_A/s,\qquad \mathsf X=D_A/t.
 \label{eq:XZ}
\end{equation}
These are Hermitian involutions with $\mathsf X\mathsf Z+\mathsf Z\mathsf X=0$, in any subspace
dimension. Work in a normalized state in this subspace, and write
\begin{equation}
 \begin{aligned}
 \alpha&=\avg{\mathsf Z\otimes U},& a&=\avg{\mathsf X\otimes \id},\\
 d&=\avg{\mathsf X\otimes B},& x&=s\alpha+td.
 \end{aligned}
 \label{eq:alphaad}
\end{equation}

\textbf{2. The anticommutation estimate.}
Let $W=\alpha \mathsf Z\otimes U+a\mathsf X\otimes \id$. The two summands anticommute,
so
\begin{equation}
 \avg W=\alpha^2+a^2,\qquad
 W^2=\alpha^2 \id\otimes U^2+a^2\id.
\end{equation}
Expanding the square gives the exact positive identity
\begin{equation}
 \begin{aligned}
 1-\alpha^2-a^2
 ={}&\avg{(\id-W)^2}\\
 &+\alpha^2\avg{\id\otimes(\id-U^2)}\geq0.
 \end{aligned}
 \label{eq:anti-witness}
\end{equation}
In particular, $\alpha^2+a^2\leq1$.

\textbf{3. Two projected means.}
Let $\Pi_\pm=(\id\pm \mathsf X)/2$ on Alice and define the unnormalized quantities
\[
 \begin{aligned}
 q_\pm&=\avg{\Pi_\pm\otimes \id}=\frac{1\pm a}{2},\\
 m_\pm&=\avg{\Pi_\pm\otimes B},\\
 (n_\pm)_i&=\avg{\Pi_\pm\otimes \mathsf R_i}.
 \end{aligned}
\]
Apply state Cauchy--Schwarz to $\Pi_\pm\otimes \id$ and $\Pi_\pm\otimes R$:
\[
 |\avg{\Pi_\pm\otimes R}|^2
 \leq q_\pm\avg{\Pi_\pm\otimes R^\dagger R}.
\]
Summing this inequality for $R=B,\mathsf R_1,\ldots,\mathsf R_{n_{\mathrm{col}}}$ and using
\eqref{eq:column} yields
\begin{equation}
 \delta_\pm:=q_\pm^2-m_\pm^2-\norm{n_\pm}^2\geq0.
 \label{eq:conditional-disks}
\end{equation}
No division by $q_\pm$ is made. If $q_\pm=0$, all its projected means
vanish automatically.

The unconditional quantities satisfy
\begin{equation}
 b=m_++m_-,\qquad d=m_+-m_-,\qquad r=n_++n_-.
\end{equation}
The parallelogram identities therefore give
\begin{equation}
 \begin{aligned}
 d^2+b^2&=2m_+^2+2m_-^2,\\
 \norm r^2&=2\norm{n_+}^2+2\norm{n_-}^2-\norm{n_+-n_-}^2.
 \end{aligned}
\end{equation}
Since $2(q_+^2+q_-^2)=1+a^2$, we obtain
\begin{equation}
 d^2+b^2+\norm r^2\leq1+a^2.
 \label{eq:d-bound}
\end{equation}
Together with \eqref{eq:anti-witness} and scalar Cauchy--Schwarz,
\[
 x^2+b^2+\norm r^2
 \leq\alpha^2+d^2+b^2+\norm r^2\leq2.
\]
The complete gap in this subspace is
\begin{equation}\begin{aligned}
 &2-x^2-b^2-\norm r^2\\
 &\quad=(1-\alpha^2-a^2)+2\delta_++2\delta_-\\
 &\qquad+\norm{n_+-n_-}^2+(t\alpha-sd)^2\geq0.
\end{aligned}\label{eq:block-gap}
\end{equation}
The first term has witness \eqref{eq:anti-witness}, and
Subsection~\ref{app:conditional-witness} gives an explicit positive
witness for $\delta_\pm$.

\textbf{4. End subspaces and averaging.}
If $t=0$ or $s=0$, one of $C_A,D_A$ vanishes and the other is an involution,
so $|x|\leq1$. Equations \eqref{eq:column-variance} and \eqref{eq:column} also give
$b^2+\norm r^2\leq1$, proving the same bound in these cases.

Let $\Pi_j$ be the spectral projections of $D_A^2$, with probabilities
$p_j=\Tr[\omega(\Pi_j\otimes \id)]>0$ for the occupied subspaces. Use the
normalized states
\begin{equation}
 \omega_j=\frac{(\Pi_j\otimes \id)\omega(\Pi_j\otimes \id)}{p_j}.
\end{equation}
All operators whose expectations define $x,b,r$ are block diagonal
on Alice. Hence, for $y_j=(x_j,b_j,r_j)$,
\begin{equation}
 y=(x,b,r)=\sum_jp_jy_j.
\end{equation}
The second subsystem and all its operators remain fixed. The exact
identity
\begin{equation}
 \begin{aligned}
 2-\norm y^2
 ={}&\sum_jp_j(2-\norm{y_j}^2)\\
 &+\sum_jp_j\norm{y_j-y}^2\geq0.
 \end{aligned}
 \label{eq:mixing-gap}
\end{equation}
proves \eqref{eq:column-bound} in arbitrary finite dimension.
Here $\norm y^2=x^2+b^2+\sum_i|r_i|^2$.
\end{proof}

This proposition assumes no source structure. Its spectral restrictions
and projected means are used only inside an unrestricted bipartite
statement. The triangle proof applies it to the full state, without
asserting independence after any conditioning on Alice.

\subsection{The independent triangle}
\label{app:triangle-physical}

Independently purify each edge, assign its purifier to one endpoint,
and extend the local observables by identities. This preserves source
independence and every correlator. It suffices to consider
\begin{equation}
 \ket\Psi=\ket\psi_{AB}\otimes\ket\varphi_{AC}\otimes\ket\chi_{BC}.
\end{equation}
Regroup the first two edges as $\ket\Omega$ on
$\mathcal H_A\otimes\mathcal R$, where
\begin{equation}
 \begin{aligned}
 \mathcal H_A&=\mathcal H_{A|B}\otimes\mathcal H_{A|C},\\
 \mathcal R&=\mathcal H_{B|A}\otimes\mathcal H_{C|A},\\
 \mathcal E&=\mathcal H_{B|C}\otimes\mathcal H_{C|B}.
 \end{aligned}
\end{equation}
Thus $\ket\Psi=\ket\Omega\otimes\ket\chi$ after regrouping, with
$\ket\Omega=\ket\psi\otimes\ket\varphi$.

\subsection{The setting-zero contraction}

Define the isometry $V:\mathcal R\to\mathcal R\otimes\mathcal E$ by
$Vv=v\otimes\ket\chi$. On $\mathcal R\otimes\mathcal E$, the physical
operator $O_0=B_0C_0$ is Hermitian because Bob's and Charlie's
observables commute. It is a contraction, with
\begin{equation}
 \id-O_0^2=(\id-B_0^2)+B_0(\id-C_0^2)B_0\succeq0.
\end{equation}
Set
\begin{equation}
 U_0=V^\dagger O_0V.
 \label{eq:triangle-U0}
\end{equation}
It is Hermitian, and its contraction property has the exact witness
\begin{equation}
\begin{aligned}
 \id-U_0^2={}&V^\dagger(\id-O_0^2)V\\
 &+V^\dagger O_0(\id-VV^\dagger)O_0V\succeq0.
 \end{aligned}
 \label{eq:triangle-U0-witness}
\end{equation}
Both summands are positive congruences. Alice's joint operators $C_A,D_A$
are unchanged.

\subsection{Operator columns and their expectation vectors}

Set $n_{\mathrm{col}}=\dim\mathcal E-1$ and choose an orthonormal
basis $\{\ket i:i=0,\ldots,n_{\mathrm{col}}\}$ of the whole
$BC$ edge space $\mathcal E$, with $\ket0=\ket\chi$. Define operators
on $\mathcal R$ by
\begin{equation}
 \begin{aligned}
 \mathsf R_i&=(\id_{\mathcal R}\otimes\bra i)B_1V,\\
 \mathsf Q_i&=(\id_{\mathcal R}\otimes\bra i)C_1V.
 \end{aligned}
 \label{eq:physical-columns}
\end{equation}
Tensor identities on endpoints not acted on are implicit. Locality gives
$\mathsf R_i=\mathsf R_i^{(B)}\otimes \id_{C|A}$ and
$\mathsf Q_i=\id_{B|A}\otimes \mathsf Q_i^{(C)}$.
Write
\begin{equation}
 B=\mathsf R_0=V^\dagger B_1V,\qquad
 C=\mathsf Q_0=V^\dagger C_1V.
\end{equation}
These are auxiliary operators on $\mathcal R$; the physical observables
retain their setting labels $B_y,C_z$. Completeness of the basis gives
\begin{equation}
 \begin{aligned}
 B^2+\sum_{i>0}\mathsf R_i^\dagger \mathsf R_i&=V^\dagger B_1^2V\preceq \id,\\
 C^2+\sum_{i>0}\mathsf Q_i^\dagger \mathsf Q_i&=V^\dagger C_1^2V\preceq \id.
 \end{aligned}\label{eq:physical-column-inequalities}
\end{equation}
Indeed, $\sum_i\mathsf R_i^\dagger \mathsf R_i=V^\dagger B_1^2V$ and $\mathsf R_0=B$ is
Hermitian. The two column deficits are explicitly
$V^\dagger(\id-B_1^2)V\succeq0$ and $V^\dagger(\id-C_1^2)V\succeq0$.
In particular, $B,C$ are Hermitian contractions.

In the bipartite state $\ket\Omega$ define
\begin{equation}
 \begin{aligned}
 x_B&=\avg{C_A\otimes U_0+D_A\otimes B}_{\Omega},\\
 x_C&=\avg{C_A\otimes U_0+D_A\otimes C}_{\Omega},\\
 z_i&=\avg{\id_A\otimes \mathsf R_i}_{\Omega},\qquad
 w_i=\avg{\id_A\otimes \mathsf Q_i}_{\Omega}.
 \end{aligned}
 \label{eq:xzw}
\end{equation}
Include the reference coefficients in
\begin{equation}
 \begin{aligned}
 v_B&=(z_0,z_1,\ldots,z_{n_{\mathrm{col}}}),\\
 v_C&=(w_0,w_1,\ldots,w_{n_{\mathrm{col}}}).
 \end{aligned}
\end{equation}
The scalars $z_0,w_0$ are real; the remaining entries may be
complex. Averaging the independent $BC$ source first yields
\begin{equation}
 \begin{aligned}
 x_B&=\avg{C_A B_0C_0+D_A B_1}_{\Psi},\\
 x_C&=\avg{C_A B_0C_0+D_A C_1}_{\Psi}.
 \end{aligned}\label{eq:physical-identities}
\end{equation}
Apply Proposition~\ref{prop:column} with $U=U_0$, first to $B,\mathsf R_{i>0}$
and then to $C,\mathsf Q_{i>0}$. Both applications use the same full state
$\ket\Omega$ and the same Alice observables. Thus
\begin{equation}
 \begin{aligned}
 \varepsilon_B&:=2-x_B^2-\norm{v_B}^2\geq0,\\
 \varepsilon_C&:=2-x_C^2-\norm{v_C}^2\geq0.
 \end{aligned}
 \label{eq:physical-tradeoffs}
\end{equation}

\subsection{The shared-source overlap and the Bell gap}

The common $BC$ basis gives
\begin{equation}
 \begin{aligned}
 B_1\ket\Psi&=\sum_i[(\id_A\otimes \mathsf R_i)\ket\Omega]\otimes\ket i,\\
 C_1\ket\Psi&=\sum_i[(\id_A\otimes \mathsf Q_i)\ket\Omega]\otimes\ket i.
 \end{aligned}
\end{equation}
Since $\mathsf R_i$ acts only on the remote $AB$ endpoint and $\mathsf Q_i$ only on
the remote $AC$ endpoint, the product
$\ket\Omega=\ket\psi\otimes\ket\varphi$ implies
\begin{equation}
 \begin{aligned}
 &\avg{\id_A\otimes \mathsf R_i^\dagger \mathsf Q_i}_{\Omega}\\
 &\quad=\avg{\id_A\otimes \mathsf R_i^\dagger}_{\Omega}
 \avg{\id_A\otimes \mathsf Q_i}_{\Omega}
 =\overline{z_i}w_i.
 \end{aligned}
\end{equation}
Therefore
\begin{equation}
 \begin{aligned}
 \avg{B_1C_1}_{\Psi}
 &=\inner{B_1\Psi}{C_1\Psi}\\
 &=\sum_i\overline{z_i}w_i=\Ree\inner{v_B}{v_C}.
 \end{aligned}
 \label{eq:physical-overlap}
\end{equation}
The sum is real because $B_1C_1$ is Hermitian. This factorization uses
independence of $AB$ and $AC$; independence of $BC$ allowed the common
isometry $V$.

Equations \eqref{eq:triangle-score} and \eqref{eq:physical-identities} now give
\begin{equation}
 I_2+1=x_B+x_C+\Ree\inner{v_B}{v_C}.
\end{equation}
Using \eqref{eq:physical-tradeoffs} and completing squares yields the
exact nonnegative gap
\begin{equation}
\begin{aligned}
 2-I_2=\frac12\bigl[&(x_B-1)^2+(x_C-1)^2\\
 &+\norm{v_B-v_C}^2+\varepsilon_B+\varepsilon_C\bigr]\geq0.
\end{aligned}\label{eq:final-gap}
\end{equation}
The two $\varepsilon$ terms are nonnegative by the block estimates
(including their endpoint cases) and the averaging identity
\eqref{eq:mixing-gap}. This proves the
finite-dimensional theorem when Alice's observables are involutions.

\subsection{Equality at every bias}
\label{app:triangle-sharpness}

Here $X,Z$ are the ordinary Pauli matrices. For $\kappa\in[0,1]$ set
\begin{equation}
 \begin{aligned}
 H_\kappa&=(1-2\kappa)Z+2\sqrt{\kappa(1-\kappa)}X,\\
 H_\kappa^2&=[(1-2\kappa)^2+4\kappa(1-\kappa)]\id=\id.
 \end{aligned}
\end{equation}
Take each independent edge state to be $\ket{00}\bra{00}$. Order the
incoming systems at $A,B,C$ as $(AB,AC)$, $(AB,BC)$, $(AC,BC)$, respectively.
For a sign $\varsigma\in\{-1,+1\}$ choose
\begin{equation}
 \begin{aligned}
 A_0&=\varsigma Z\otimes \id,\qquad A_1=\varsigma H_\kappa\otimes \id,\\
 B_0&=B_1=\varsigma Z\otimes \id,\\
 C_0&=Z\otimes \id,\qquad C_1=\varsigma Z\otimes \id.
 \end{aligned}\label{eq:sharp-settings}
\end{equation}
All these observables are reflections, with outcome-$\pm1$
projectors $(1\pm O)/2$.
Direct calculation gives
\begin{equation}
 \begin{aligned}
 \mu&=\frac{\varsigma-\varsigma(1-2\kappa)}2=\varsigma\kappa,\\
 I_2&=2(1-\kappa)+2\kappa+1-1=2.
 \end{aligned}
\end{equation}
For any prescribed $\mu\in[-1,1]$, take $\kappa=|\mu|$ and choose $\varsigma \kappa=\mu$;
either sign works at zero. Each bias is attained by one independent-source
realization. In \eqref{eq:final-gap}, this construction has
$x_B=x_C=1$ and $v_B=v_C=(\varsigma,0,\ldots,0)$, so every gap term vanishes.

\subsection{General measurements and infinite dimensions}
\label{app:triangle-extensions}

\paragraph{Alice's two-outcome measurements.}
For general Alice effects $A_{\pm|x}$ from \eqref{eq:effects},
$A_x=A_{+|x}-A_{-|x}=1-2A_{-|x}$ is a Hermitian contraction.
On one common local qubit
ancilla, initialized in $\ket0$, dilate $A_x\to\widehat A_x$ as in
\eqref{eq:dilation}.

The square root commutes with $A_x$ by spectral calculus, so block
multiplication proves the involution identity. Compression to
$\ket0$ gives $A_x$ for both settings. Bob's and Charlie's observables
already satisfy the hypotheses of Appendix~\ref{app:triangle-physical}, so no
additional dilation is needed there.

Every expanded Bell correlator and each bias mean contains at most one
setting from each party. Consequently the common Alice dilation preserves
all of them. Assign its pure ancilla to Alice's endpoint of the $AB$
source; the three source states remain independent. Applying the result
already proved to $\widehat A_0,\widehat A_1$ establishes the finite theorem
for arbitrary two-outcome measurements at all three parties.

\paragraph{Normal infinite-dimensional sources.}
Purify each source independently and assign each purifier to an endpoint.
A pure bipartite vector has normalized finite Schmidt truncations
converging in Hilbert-space norm. Choose such a truncation on each of the
three edges. Their product $\ket{\Psi_n}$ remains a product of independent
sources and obeys $\norm{\Psi_n-\Psi}\to0$.

Compress each local observable to the finite product of the corresponding
endpoint supports. These compressions are Hermitian contractions, so the
finite theorem just proved applies. Their Bell correlators equal those
of the uncompressed observables in $\ket{\Psi_n}$, since each expanded
term contains at most one operator per party.

For unit vectors $\psi,\phi$ and a contraction $O$,
\begin{equation}
 \begin{aligned}
 &|\inner{\psi}{O\psi}-\inner{\phi}{O\phi}|\\
 &\quad\leq|\inner{\psi-\phi}{O\psi}|+|\inner{\phi}{O(\psi-\phi)}|\\
 &\quad\leq2\norm{\psi-\phi}.
 \end{aligned}
\end{equation}
Apply this estimate to each expanded Bell correlator and take the limit.
This proves $I_2\leq2$ for normal infinite-dimensional source states and
arbitrary bounded binary observables, completing Theorem~\ref{thm:triangle-bias}.

\subsection{A positive witness for the projected-mean deficits}
\label{app:conditional-witness}

In the proof of Proposition~\ref{prop:column}, let $\omega$ be the
normalized state in the spectral subspace under consideration and set
\begin{equation}
 \tau_\pm=\Tr_A[(\Pi_\pm\otimes \id)\omega(\Pi_\pm\otimes \id)]\succeq0.
\end{equation}
These unnormalized operators have trace $q_\pm$ and give the projected
means $m_\pm,n_\pm$. For either sign write $\tau=\tau_\pm$, and put
$L_0=B$, $L_i=\mathsf R_i$ for $i>0$. Define
\begin{equation}
 \mathcal D_i=L_i\otimes\id-\id\otimes L_i.
\end{equation}
Since $D_{\rm col}=\id-\sum_{i=0}^{n_{\mathrm{col}}}L_i^\dagger L_i$,
direct expansion gives
\begin{equation}
\begin{aligned}
 \delta_\pm={}&\Tr[(\tau\otimes\tau)(D_{\rm col}\otimes\id)]\\
 &+\frac12\sum_{i=0}^{n_{\mathrm{col}}}
 \Tr[(\tau\otimes\tau)\mathcal D_i^\dagger\mathcal D_i]\geq0.
\end{aligned}\label{eq:disk-witness}
\end{equation}
The first term is
$q_\pm^2-q_\pm\sum_i\Tr(\tau L_i^\dagger L_i)$, while the sum of
squares contributes
$q_\pm\sum_i\Tr(\tau L_i^\dagger L_i)-\sum_i|\Tr(\tau L_i)|^2$.
Their sum is exactly \eqref{eq:conditional-disks}.

In each trace, both operator factors are positive. The trace of a product
of two positive operators is nonnegative, since it equals the trace of
a positive congruence. No commutation between the $L_i$, or between
$\tau$ and the $L_i$, is assumed. If $q_\pm=0$, positivity forces
$\tau=0$, and both sides vanish.

\section{Native circuit mapping and polynomial conventions}\label{app:ibm-gates}
\subsection{Logical preparation and the measurement frame}
All rotations use $G_\vartheta=\exp(-\ii\vartheta G/2)$, with
$G_0=1$ and $G_\pm=G_{\pm\pi/2}$; $G_\pi\equiv G$ up to global phase.
Here $G$ is a Pauli matrix from Appendix~\ref{app:notation}. With $H$ the
Hadamard gate, the following identities hold up to global phase:
\begin{align}\label{eq:native-identities}
 H&=Z_+X_+Z_+,\notag\\
 X_-&=ZX_+Z,\notag\\
 CX&=(1\otimes H)CZ(1\otimes H).
\end{align}
Here $CZ\ket{ab}=(-1)^{ab}\ket{ab}$ and
$CX\ket{ab}=\ket{a,a\oplus b}$, where $\oplus$ denotes addition modulo $2$.
Products in ordinary operator equations act from right to left.
For example, the measurement unitary is $V_\alpha=X_+ Z_{\pi/2-\alpha}$,
not the reverse product. Direct conjugation gives
\begin{equation}
 V_\alpha^\dagger ZV_\alpha
 =Z_{\alpha-\pi/2}Y Z_{\pi/2-\alpha}
 =X\cos\alpha+Y\sin\alpha.
\end{equation}
The $F=X_-$ frame then converts the hardware $XY$ settings into the
paper's $XZ$ four-context observables. The hardware state is locally
unitarily equivalent to the phase-zero GHZ state; local-unitary invariance
preserves the entanglement classification and its biseparable bounds.

\begin{figure*}[tbp]
\centering
\begin{tikzpicture}
\begin{yquant}
 qubit {$\ket0_{91}$} r;
 qubit {$\ket0_{98}$} b;
 qubit {$\ket0_{111}$} a;
 h r;
 cnot b | r;
 cnot a | b;
 box {$Z_\phi$} r;
 box {$U$} r,b,a;
 barrier r,b,a;
 box {$Z_{\pi/2-\alpha_j}$} r,b,a;
 box {$X_+$} r,b,a;
 measure r,b,a;
\end{yquant}
\end{tikzpicture}
\caption{Representative three-qubit circuit in \texttt{yquant}. The wire
labels are actual physical qubits; the logical order is $(91,98,111)$.
For four-context measurements take $\phi=0$, $U=F=X_-$,
$\alpha_r=\alpha_b=y\pi/2$, and $\alpha_a=(-1)^x\pi/4$.
For MABK or Svetlichny take $U=1$, $\alpha_j=s_j\pi/2$, and the specified
study phase. Identity boxes are omitted in the actual circuit. Hadamards
and CNOTs are symbolic abbreviations for \eqref{eq:native-identities};
the saved circuits contain only native gates. Fixed Clifford folding may
shorten the preparation but cannot move a setting-dependent operation
across the barrier. The barrier is a circuit boundary, not a physical
spacetime-isolation guarantee.}
\label{fig:yquant-ibm}
\end{figure*}
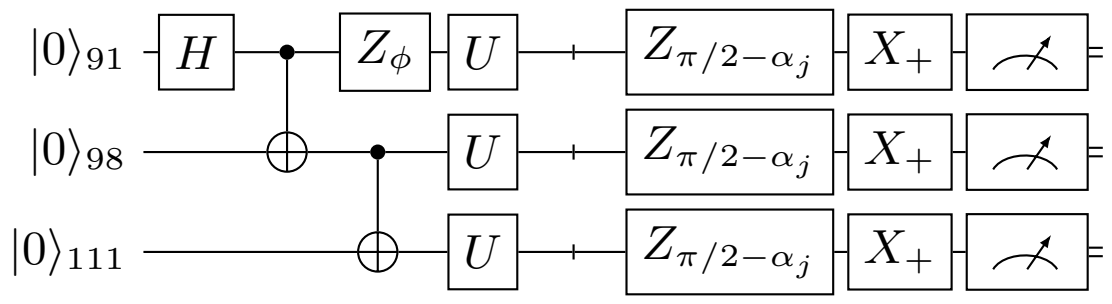

For all $N$, preparing $H$ on the root and then the recorded directed tree
CNOTs gives \eqref{eq:ghzdef} with $\phi=0$. Applying $R_z(\phi)$ to the
root inserts the relative phase $\phi$. The ideal full-parity prediction
for hardware setting mask $s=(s_0,\ldots,s_{N-1})$ is
\begin{equation}\label{eq:hardware-parity}
 E_s=\cos\!\left(\phi-\frac\pi2|s|\right),\qquad
 |s|=\sum_j s_j.
\end{equation}
This fixes both the phase sign and the binary-input convention without
fitting the hardware data.

\subsection{Integer MABK scores and the paper's normalization}
The archived code evaluates the integer-coefficient polynomial
\begin{align}\label{eq:raw-mabk}
 \widetilde M_N&=\sum_s c_N(|s|)E_s,\notag\\
 c_N(w)&=\frac{\Ree[(1+\ii)^{N-1}\ii^w]}
 {2^{\lfloor(N-1)/2\rfloor}}\in\{0,\pm1\}.
\end{align}
Its prescribed state phase is $\phi_{\rm M}=(1-N)\pi/4$ modulo $2\pi$.
The stored native sequence wraps this angle into $[-\pi,\pi)$ but does
not optimize it. The code's local bound is $2^{\lfloor N/2\rfloor}$ and
its quantum maximum is that number times $2^{(N-1)/2}$.

To obtain exactly \eqref{eq:mabk}, relabel the output sign on input $1$
at every party: $A'_{j0}=X_j$, $A'_{j1}=-Y_j$. Then $E'_s=(-1)^{|s|}E_s$
and
\begin{equation}\label{eq:mabk-code-map}
 \mathcal B_N(A')=\frac{\widetilde M_N}{2^{\lfloor N/2\rfloor}}.
\end{equation}
This deterministic relabeling reverses the complex-product convention;
it is not a maximization over data. Equation~\eqref{eq:ibm-scale} is
therefore exactly $2\sqrt2$ times the archived score divided by its
quantum maximum. An error bar is multiplied by the same positive factor.

\subsection{Svetlichny and the even-size reuse}
Let $\mathcal M_N=\prod_j(A_{j0}+\ii A_{j1})$ as in the main text. On
one specified party, chosen here as the preparation root, use the fixed
quarter-turn relabeling $T_k$:
\begin{equation}\label{eq:quarter-relabel}
\begin{array}{c|c}
 k\pmod4&(A'_{0},A'_{1})\\\hline
 0&(A_0,A_1)\\
 1&(-A_1,A_0)\\
 2&(-A_0,-A_1)\\
 3&(A_1,-A_0).
\end{array}
\end{equation}
It multiplies the complex product by $\ii^k$. Thus it combines an input
permutation and prescribed output signs, all independent of the data.

For the dedicated odd-$N$ study, the coefficient prefactor is
$\sqrt2 e^{\ii(N-2)\pi/4}=(1+\ii)\ii^{(N-3)/2}$. Applying
$T_{(N-3)/2}$ identifies the raw polynomial with
$\mathcal S_N=\Ree[(1+\ii)\avg{\mathcal M_N}]$. The native phase is
$\phi_{\rm S}=(2-N)\pi/4$ modulo $2\pi$, corresponding to the canonical
phase $-\pi/4$ after this relabeling. The values of $k$ at $N=3,5,7,9$
are $0,1,2,3$, respectively.

For even $N$, the MABK integer prefactor instead obeys
\begin{equation}
 \frac{(1+\ii)^{N-1}}{2^{(N-2)/2}}
 =(1+\ii)\ii^{(N-2)/2}.
\end{equation}
Consequently $T_{(N-2)/2}$ turns the \emph{original hardware} MABK
polynomial directly into the paper's Svetlichny polynomial. At
$N=4,6,8,10$, use $k=1,2,3,0$ modulo $4$. This is an alternative mapping
of the original counts, not an additional relabeling to be composed with
\eqref{eq:mabk-code-map}. The normalizations give the exact identity
$W_N^{\rm S}=W_N^{\rm M}$ for these reused data, including their standard
errors and all underlying sample fluctuations.

\subsection{Sufficient statistics and reproducibility checks}
For every circuit, the compressed CSV stores an outcome string and its
integer multiplicity. Summing multiplicities recovers the declared shot
count; missing rows mean zero observations, not discarded outcomes. For
a four-context circuit the sufficient moments are
\begin{equation}
 (n,\ \sum t,\ \sum t^2,\ \sum c,\ \sum tc).
\end{equation}
For $y=1$, these give
\begin{align}
 \sum f_\lambda&=\sum t-\frac\lambda2(n-\sum c),\notag\\
 \sum f_\lambda^2&=\sum t^2+\lambda(\sum tc-\sum t)
 +\frac{\lambda^2}{4}(n-\sum c).
\end{align}
Thus all penalty curves and their covariance-aware uncertainties follow
from the same four contexts. For any per-circuit score $f$, the unbiased
independent-shot estimate of the variance of its sample mean
$\bar f=n^{-1}\sum f$ is
\begin{equation}
 \widehat{\Var}(\bar f)
 =\frac{\sum f^2-(\sum f)^2/n}{n(n-1)}.
\end{equation}
The release's independent analysis uses this expression and the fixed
shot weights in \eqref{eq:variance}. It also retains pooled shot errors
and submission-sweep diagnostics, without treating either as an
adversarial-memory guarantee.


\begin{thebibliography}{99}
\bibitem{bell} J.~S. Bell, On the Einstein Podolsky Rosen paradox,
Physics Physique Fizika \textbf{1}, 195--200 (1964),
\doiurl{10.1103/PhysicsPhysiqueFizika.1.195}.
\bibitem{chsh} J.~F. Clauser, M.~A. Horne, A. Shimony, and R.~A. Holt,
Proposed experiment to test local hidden-variable theories,
Phys. Rev. Lett. \textbf{23}, 880--884 (1969),
\doiurl{10.1103/PhysRevLett.23.880}.
\bibitem{Brunner2014} N. Brunner, D. Cavalcanti, S. Pironio, V. Scarani,
and S. Wehner, Bell nonlocality,
Rev. Mod. Phys. \textbf{86}, 419--478 (2014),
\doiurl{10.1103/RevModPhys.86.419}.
\bibitem{mermin} N.~D. Mermin, Extreme quantum entanglement in a
superposition of macroscopically distinct states,
Phys. Rev. Lett. \textbf{65}, 1838 (1990),
\doiurl{10.1103/PhysRevLett.65.1838}.
\bibitem{ardehali} M. Ardehali, Bell inequalities with a magnitude of
violation that grows exponentially with the number of particles,
Phys. Rev. A \textbf{46}, 5375 (1992),
\doiurl{10.1103/PhysRevA.46.5375}.
\bibitem{beli} A.~V. Belinskii and D.~N. Klyshko,
Interference of light and Bell's theorem,
Phys. Usp. \textbf{36}, 653 (1993),
\doiurl{10.1070/PU1993v036n08ABEH002299}.
\bibitem{tsirel} B.~S. Cirel'son, Quantum generalizations of Bell's
inequality, Lett. Math. Phys. \textbf{4}, 93--100 (1980),
\doiurl{10.1007/BF00417500}.
\bibitem{ghz89}
D. M. Greenberger, M.A. Horne, A. Zeilinger,  Going Beyond Bell's Theorem, in: M. Kafatos (ed.), Bell's Theorem, Quantum Theory and Conceptions of the Univers.e Fundamental Theories of Physics, vol 37, p. 69. Springer, Dordrecht (1989),
\doiurl{10.1007/978-94-017-0849-4_10}.
\bibitem{ghz90}
D. M. Greenberger, M.A. Horne, A. Shimony, A. Zeilinger,
 Bell's theorem without inequalities, Am. J. Phys. 58, 1131 (1990)
 \doiurl{10.1119/1.16243}.
\bibitem{seevinck} M. Seevinck and G. Svetlichny,
Bell-type inequalities for partial separability in $N$-particle systems
and quantum mechanical violations,
Phys. Rev. Lett. \textbf{89}, 060401 (2002),
\doiurl{10.1103/PhysRevLett.89.060401}.
\bibitem{seevinck01}
M. Seevinck,
J. Uffink,
Sufficient conditions for three-particle entanglement and their tests in recent experiments,
Phys. Rev. A 65, 012107 (2001)
\doiurl{10.1103/PhysRevA.65.012107}.
\bibitem{nagata}
K. Nagata, M. Koashi, and N. Imoto,
Configuration of Separability and Tests for Multipartite Entanglement in Bell-Type Experiments,
Phys. Rev. Lett. \textbf{89}, 260401 (2002)
\doiurl{10.1103/PhysRevLett.89.260401}.
\bibitem{uff} J. Uffink, Quadratic Bell inequalities as tests for
multipartite entanglement,
Phys. Rev. Lett. \textbf{88}, 230406 (2002),
\doiurl{10.1103/PhysRevLett.88.230406}.
\bibitem{svetlichny} G. Svetlichny,
Distinguishing three-body from two-body nonseparability by a Bell-type
inequality, Phys. Rev. D \textbf{35}, 3066 (1987),
\doiurl{10.1103/PhysRevD.35.3066}.
\bibitem{bancal11}
J.-D. Bancal, N. Brunner, N. Gisin, and Y.-C. Liang,
Detecting Genuine Multipartite Quantum Nonlocality: A Simple Approach and Generalization to Arbitrary Dimensions,
Phys. Rev. Lett. \textbf{106}, 020405 (2011).
\doiurl{10.1103/PhysRevLett.106.020405}.
\bibitem{collinsm4} D. Collins, N. Gisin, S. Popescu, D. Roberts, and
V. Scarani, Bell-type inequalities to detect true $n$-body nonseparability,
Phys. Rev. Lett. \textbf{88}, 170405 (2002),
\doiurl{10.1103/PhysRevLett.88.170405}.
\bibitem{Bancal2013} J.-D. Bancal, J. Barrett, N. Gisin, and S. Pironio,
Definitions of multipartite nonlocality,
Phys. Rev. A \textbf{88}, 014102 (2013),
\doiurl{10.1103/PhysRevA.88.014102}.
\bibitem{coiteux}
X. Coiteux-Roy, E. Wolfe, and M.-O. Renou,
No Bipartite-Nonlocal Causal Theory Can Explain Nature’s Correlations,
Phys. Rev. Lett. \textbf{127}, 200401 (2021)
\doiurl{10.1103/PhysRevLett.127.200401}.
\bibitem{coiteux2}
X. Coiteux-Roy, E. Wolfe, and M.-O. Renou,
Any Physical Theory of Nature Must Be Boundlessly Multipartite Nonlocal,
Phys. Rev. A \textbf{104}, 052207 (2021).
\doiurl{10.1103/PhysRevA.104.052207}.
\bibitem{cao}
H. Cao \emph{et al.},
Experimental Demonstration that No Tripartite-Nonlocal Causal Theory Explains Nature’s Correlations,
Phys. Rev. Lett. \textbf{129}, 150402 (2022),
\doiurl{10.1103/PhysRevLett.129.150402}.
\bibitem{mao}
Y.-L. Mao, Z.-D. Li, S. Yu, and J. Fan,
Test of Genuine Multipartite Nonlocality,
Phys. Rev. Lett. \textbf{129}, 150401 (2022)
\doiurl{10.1103/PhysRevLett.129.150401}.
\bibitem{suprano}
A. Suprano, D. Poderini, E. Polino, I. Agresti, G. Carvacho, A. Canabarro, E. Wolfe, R. Chaves, and 
F. Sciarrino,
Experimental Genuine Tripartite Nonlocality in a Quantum Triangle Network,
PRX Quantum \textbf{3}, 030342 (2022),
\doiurl{10.1103/PRXQuantum.3.030342}.
\bibitem{navasques}
M. Navascues, E. Wolfe, D. Rosset, and A. Pozas-Kerstjens,
Genuine Network Multipartite Entanglement,
Phys. Rev. Lett. 125, 240505 (2020)
\doiurl{10.1103/PhysRevLett.125.240505}.
\bibitem{wolfe}
E. Wolfe, A. Pozas-Kerstjens, M. Grinberg, D. Rosset, A. Acin, and M. Navascues,
Quantum Inflation: A General Approach to Quantum Causal Compatibility,
Phys. Rev. X \textbf{11}, 021043 (2021)
\doiurl{10.1103/PhysRevX.11.021043}.
\bibitem{Contreras2021} P. Contreras-Tejada, C. Palazuelos, and
J.~I. de Vicente, Genuine multipartite nonlocality is intrinsic to
quantum networks, Phys. Rev. Lett. \textbf{126}, 040501 (2021),
\doiurl{10.1103/PhysRevLett.126.040501}.
\bibitem{Finner1992} H. Finner, A generalization of H\"older's inequality
and some probability inequalities,
Ann. Probab. \textbf{20}, 1893--1901 (1992),
\doiurl{10.1214/aop/1176989534}.
\bibitem{Renou2019} M.-O. Renou, Y. Wang, S. Boreiri, S. Beigi,
N. Gisin, and N. Brunner, Limits on correlations in networks for quantum
and no-signaling resources,
Phys. Rev. Lett. \textbf{123}, 070403 (2019),
\doiurl{10.1103/PhysRevLett.123.070403}.
\bibitem{barreiro}
J. T. Barreiro, J.-D. Bancal, P. Schindler, D. Nigg, M. Hennrich, T. Monz, N. Gisin, and R. Blatt,
Demonstration of genuine multipartite entanglement with device-independent witnesses,
Nature Physics 9, 559 (2013),
\doiurl{10.1038/nphys2705}.
\bibitem{wang}
K. Wang \emph{et al.},
Probing many-body Bell correlation depth with superconducting qubits,
 	Physical Review X 15, 021024 (2025),
\doiurl{10.1103/PhysRevX.15.021024}.
\bibitem{swain}
M. Swain, A. Rai, B. K. Behera, P. K. Panigrahi,
Experimental demonstration of the violations of Mermin's and Svetlichny's inequalities for W- and GHZ-class of states,
Quantum Inf. Process. 18, 218 (2019)
\doiurl{10.1007/s11128-019-2331-5}.
\bibitem{alsina}
 D. Alsina and J. I. Latorre, 
Experimental test of Mermin inequalities on a five-qubit quantum computer,
Phys. Rev. A 94, 012314
(2016),
\doiurl{10.1103/PhysRevA.94.012314}.
\bibitem{gonzalez}
D. Gonzalez, D. F. de la Pradilla, G. Gonzalez,
Revisiting the experimental test of Mermin's inequalities at IBMQ,
 	Int. J. Theor. Phys. 59, 3756 (2020)
 (2020)
\doiurl{10.1007/s10773-020-04629-4}.
\bibitem{zen}
Supplementary data and numerical materials for this work, Zenodo,
\href{https://doi.org/10.5281/zenodo.24080024}{\nolinkurl{doi:10.5281/zenodo.24080024}}.
\bibitem{baccari}
F. Baccari, D. Cavalcanti, P. Wittek, and A. Acin,
Efficient Device-Independent Entanglement Detection for Multipartite Systems,
Phys. Rev. X \textbf{7}, 021042 (2017),
\doiurl{10.1103/PhysRevX.7.021042}.
\bibitem{baccari2}
F. Baccari, R. Augusiak, I. Supic, J. Tura, and A. Acin,
Scalable Bell Inequalities for Qubit Graph States and Robust Self-Testing,
Phys. Rev. Lett. \textbf{124}, 020402 (2020).
\doiurl{10.1103/PhysRevLett.124.020402}.
\bibitem{almeida}
M. L. Almeida, J.-D. Bancal, N. Brunner, A. Acin, N. Gisin, 
Guess Your Neighbor’s Input: A Multipartite Nonlocal Game with No Quantum Advantage
and S. Pironio, Phys. Rev. Lett. 104, 230404 (2010),
\doiurl{10.1103/PhysRevLett.104.230404}.
\bibitem{kofler}
J. Kofler, M. Giustina, J,-A. Larsson, and M. W. Mitchell,
Requirements for a loophole-free photonic Bell test using imperfect setting generators,
Phys. Rev. A 93, 032115 (2016),
\doiurl{10.1103/PhysRevA.93.032115}.
\bibitem{pironio} S. Pironio \emph{et al.},
Random numbers certified by Bell's theorem,
Nature \textbf{464}, 1021--1024 (2010),
\doiurl{10.1038/nature09008}.
\bibitem{Braunstein1990} S.~L. Braunstein and C.~M. Caves,
Wringing out better Bell inequalities,
Ann. Phys. \textbf{202}, 22--56 (1990),
\doiurl{10.1016/0003-4916(90)90339-P}.
\bibitem{wehner} S. Wehner, Tsirelson bounds for generalized
Clauser-Horne-Shimony-Holt inequalities,
Phys. Rev. A \textbf{73}, 022110 (2006),
\doiurl{10.1103/PhysRevA.73.022110}.
\bibitem{ibm} IBM Quantum,
\href{https://quantum.cloud.ibm.com}{IBM Quantum},
IBM Quantum documentation.

\bibitem{Zhang2011} Y. Zhang, S. Glancy, and E. Knill,
Asymptotically optimal data analysis for rejecting local realism,
Phys. Rev. A \textbf{84}, 062118 (2011),
\doiurl{10.1103/PhysRevA.84.062118}.
\bibitem{WignerYanase1963} E.~P. Wigner and M.~M. Yanase,
Information contents of distributions,
Proc. Natl. Acad. Sci. USA \textbf{49}, 910--918 (1963),
\doiurl{10.1073/pnas.49.6.910}.
\end{thebibliography}
\end{document}